\documentclass[a4paper,12pt]{article}
\usepackage{graphicx}
\usepackage{amsfonts}
\usepackage{amsmath}

\author{Liubov Tupikina}

\title{Dynamics on networks}

\begin{document}
\maketitle
\section{Introduction} 
\label{ch1}
\begin{figure*}[h]
\includegraphics[width=0.55 \textwidth]{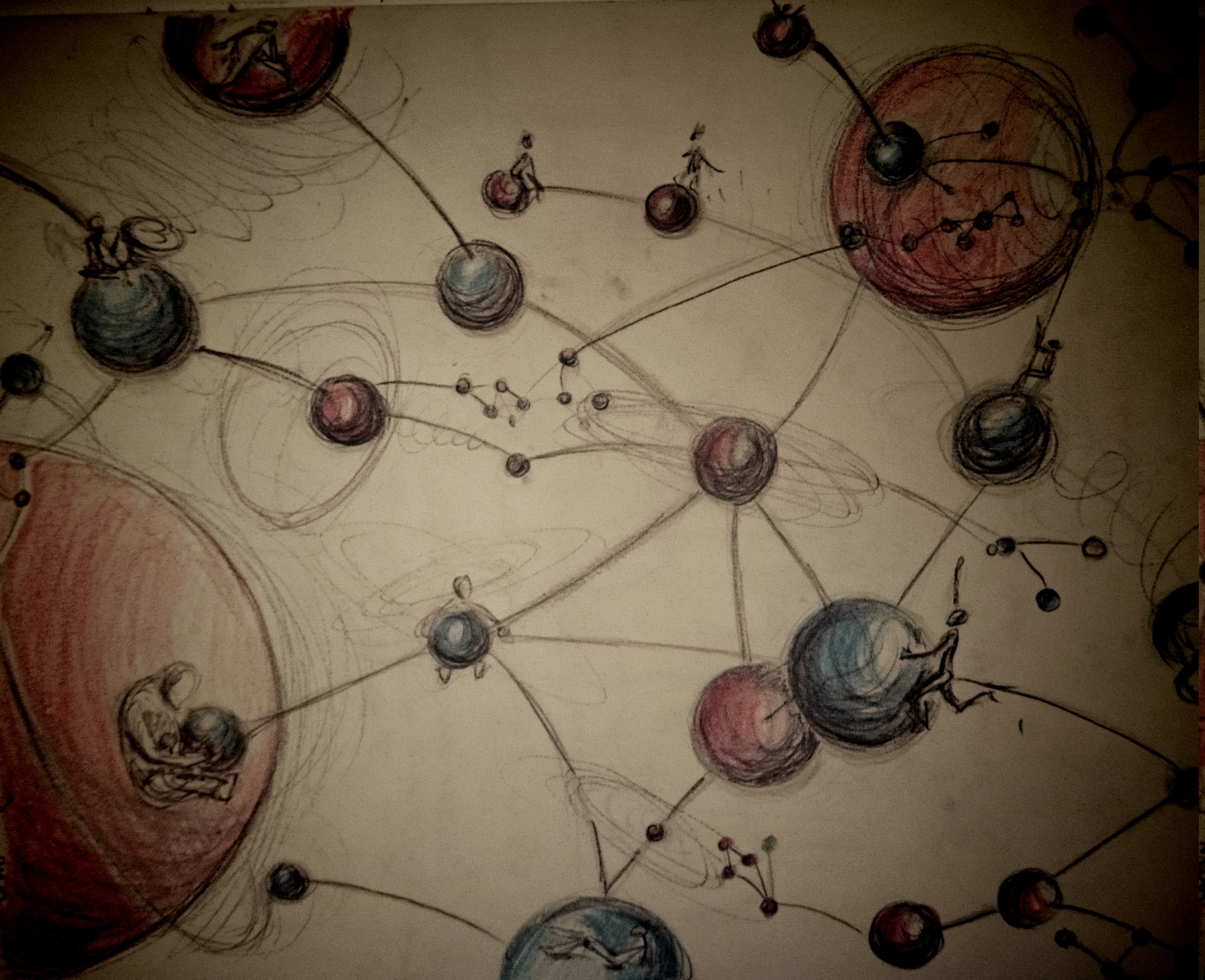} \centering
\caption{\textbf{Illustrative example of a dynamical network model}: 
nodes represent humans (or separate communities). 
Links between nodes correspond to connections between humans, 
nodes size represents status of a human (or his/her importance in society) in the HOpS model. 
Links 
and states of the nodes can evolve in time.} 
\label{discre_stat}
\end{figure*}

.
Recently an extensive and detailed graph theoretical analysis of networks with applications 
to neurobiology, climate and power grids 
has been performed, 
and has been particularly discussed in Chapters II and III. 
A particular example of a complex system is the Earth evolution which 
cannot be described without a "human
factor" anymore \cite{Schellnhuber, ramstorf2015}.
Such a system needs to be considered in coexistence with other components. 
Recently the concept of 
planetary boundaries \cite{Steffen2015a} has been introduced, 
where different components of the Earth system are considered together
in co-called \emph{co-evolution}. 
Co-evolutionary modeling approaches aim at 
incorporating the complex dynamics of society into the 
description of natural systems in order to obtain a more holistic 
picture of the world-earth system. 
As our world becomes increasingly connected through the use of 
communication and transportation systems, 
an understanding of how these connecting networks evolve in time 
plays an important role. 
\\
As an attempt to understand some mechanisms of the complex systems,  
models on networks with dynamically changing parameters (\emph{graph dynamical systems} or \emph{dynamical network models}) 
have been mathematically described in \cite{Macauley2000} and later on further designed in \cite{Arenas2008, Lambiotte2010}. 
The nodes of a \emph{dynamical network} (DN) are individual dynamical systems which are 
coupled through static links. Moreover the network topology can evolve dynamically in time. 
As the result, combination of dynamics \emph{on} networks and dynamics \emph{of} networks 
yields a 
particular class of the dynamical networks,   
so-called  \emph{adaptive network models} \cite{Gross2009}. 
Another class of dynamical networks are \emph{discrete state network models}, 
where a state of each node is defined by a discrete function evolving in time.   
An illustrative example of such a model, where each node has a discrete state, 
is shown in 
Fig.~\ref{discre_stat}.
Studies of analytical and numerical solutions for DN models 
become a topical issue in natural science \cite{Sayama2014, Auer}. 
DN models have been successfully studied using graph theoretical approaches, 
algebraic groups properties of graphs, 
probability theory and Markov chains \cite{Harary1969, Weber2012}. 
One such approach was explicitly 
demonstrated in \cite{freidlin}, 
where a network is defined by the transformation matrix of 
a  Markov chain. 
In particular, directed graphs can be interpreted in the sense 
that events are represented by nodes of the graph, 
and a directed line from one node to another indicates 
a positive probability of direct succession of these two events. 
Several of the concepts listed above have been successfully applied 
to describe a broad spectrum of various types of DN. 
However it is hard to develop 
a general theoretical framework for investigation of 
analytical solutions for DN models 
since they have structural differences. 

Here I developed a new conceptual, stochastic 
\textbf{H}eterogeneous \textbf{Op}inion-\textbf{S}tatus 
model (HOpS model), presented in details in Section \ref{dyn_netw_mod}.  
The \emph{HOpS model} admits to identify 
the main attributes of dynamics on networks    
and to study analytically the relation between topological network properties and processes 
taking place on a network. 
Another key point 
of the \emph{HOpS model} is the possibility to study network dynamics 
via the novel parameter of heterogeneity. 
I show that not only clear topological network properties, such as node degree, but also 
the nodes' status distribution 
play an important 
role in so-called opinion spreading and information diffusion on a network, Subsection \ref{model_set}. 
Furthermore, in Section \ref{res_hops} 
I propose an analytical method to study DN models 
demonstrating it on the \emph{HOpS model} on networks with regular topologies. 
The analytic solutions   
are also extended by the numerical results from Subsection \ref{num_res}.  

\subsection{Motivation}
\label{res_ques}

\begin{figure*}[h]
\includegraphics[width=0.78 \textwidth]{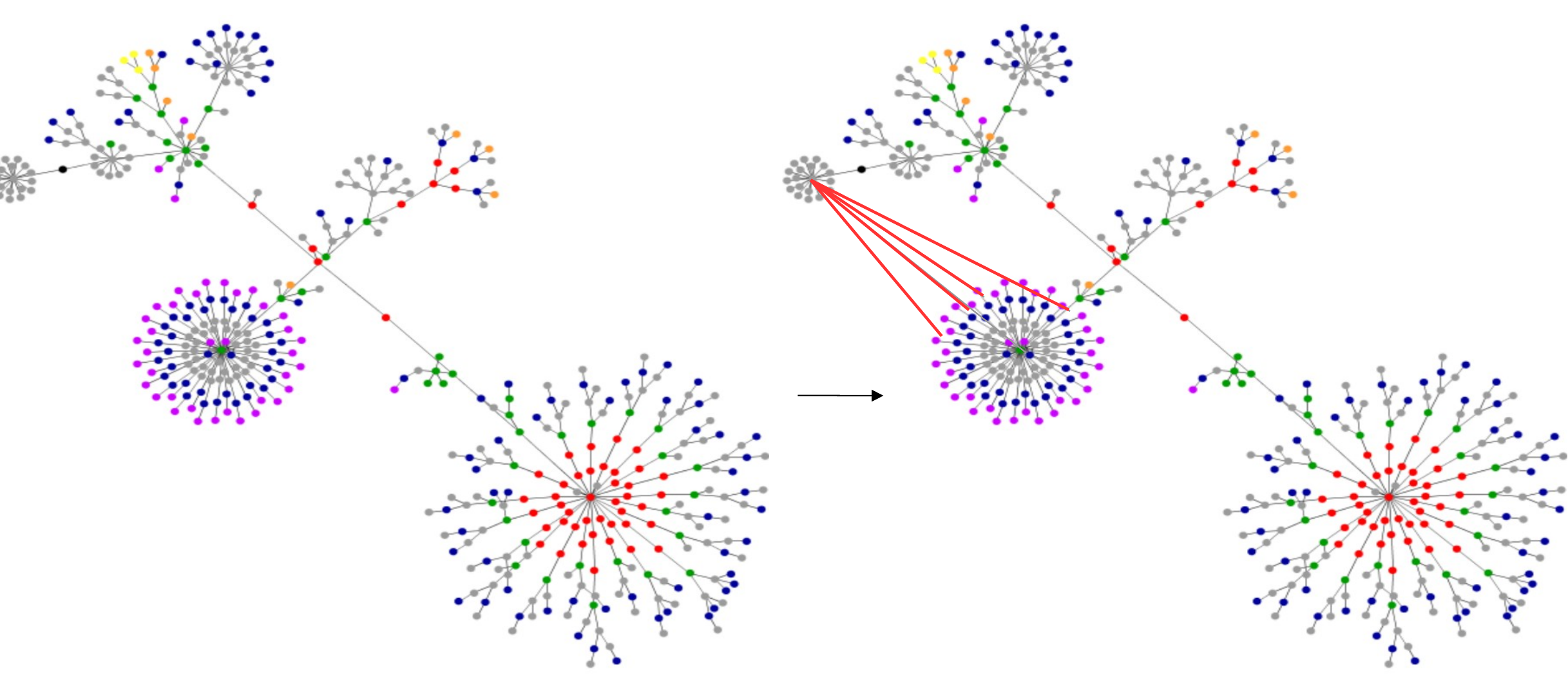} \centering 
\caption{
{\bf Evolving twitter-network} is shown in consequent time steps: at time step $t$ (a), 
{and at time step $t+1$} (b), when several  
new connections are emerged in the network, highlighted in red.}
\label{interaction1}
\end{figure*}

\begin{figure*}[h]
\includegraphics[width=0.6 \textwidth]{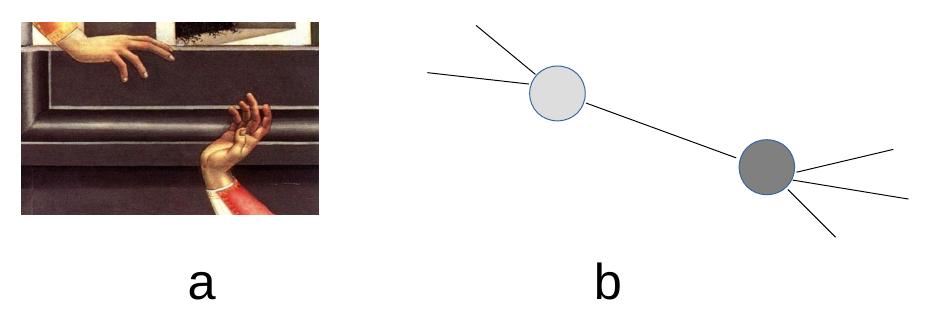} \centering 
\caption{
{\bf Connections between people} have been topic 
of discussions besides artists and scientists since the ancient times. 
The fragment from Botticelli's painting (a)  illustrates a human interaction (b). 
}
\label{inter}
\end{figure*}

The process of "spreading out" of a substance is widely used in physics (particle diffusion), 
chemistry, sociology and others \cite{sokolov2012, Metzler}. 
\emph{Diffusion} is a fundamental transport mechanism with countless examples in nature 
\cite{gomez2012, Tamm2014, Colliza}, 
which leave many open fundamental questions \cite{barkai2012, coifman2006}. 
The molecular nature of homogeneous diffusion was understood 
using new approach of Einstein to a random walk 
\cite{einstein1905}. The study of random walks on different structures 
such as regular lattices or, for instance, Cayley graphs \cite{Kohler1990}, 
allows to understand 
how certain dynamical processes on networks take place,  
for instance, energy transfer, chemical reactions and 
transport problems.
Moreover a variety of interesting mathematical problems arise from these studies \cite{Shiryaev2012}. 
Spatial aspects of diffusion and advection processes 
were recently studied using the flow-networks approach presented in \cite{rossi2014, Tupikina2015, Kutza2015a},
and were discussed in details in Chapter III. 
Flow-networks are constructed from a discretisation of 
the advection-diffusion equation on regular grids, 
which helps to bridge the gap between the dynamics of the system and the topology of 
the corresponding correlation network. 
While the purpose of \emph{functional correlation networks} (FCN)  
 is to study data time-series or a dynamical systems from 
obtained topological properties of FCN,   
the purpose of so-called 
\emph{dynamical networks} is to study processes on networks with a "prescribed" topology, 
which can be in addition coupled with dynamics on a network \cite{hueve}. 
It is clear 
that the combination (or in other words, adaptation) of non-trivial network topologies and dynamical processes
on a network can produce rich dynamics.  
In many recent works on opinion and coalition formation \cite{Gross2007, Sayama2014, Auer} 
\emph{dynamical adaptive networks} were used as a prominent tool to analyze complex systems. 
One can define a variety of DN models  
on less regular networks, such as 
small-world networks and many others.  
Several statistical physics concepts were introduced 
to describe adaptive dynamics \cite{castellano2009}, which can also 
be applied to study social collective behavior. %
It is not necessary to justify that opinion formation processes 
play an important role in many aspects of our life \cite{heitzig2011, Hegselmann2002}. 
The last years have seen a clear rise of interest in collective phenomena emerging 
from the interactions between individuals in social structures. 
Typically, 
society structure is represented as a network 
in mathematical approaches to this problem,
where a link determines the connection between nodes, see Fig.~\ref{inter}. 
The ubiquitous real-world examples demonstrate, why it is important to model  
the information spread processes using DN models defined on networks:  
\\
\emph{Example 1.}
Person $T$ lives in country $A$. 
$T$ is interacting every day with many people, but only 
from country $A$. 
After some time person $T$ gets a letter from another 
person from country $B$ 
with some information about himself (about person $T$). 
How is this possible? The reason is that some friend of person $T$, 
living in country $A$, 
traveled to country $B$  and spread the information 
about person $T$ to people from country $B$. 
\\
\emph{Example 2.} 
One user with a few connections in some internet network 
wrote some news which were highlighted ("liked") 
by some "big hub" user in this social network. 
As a consequence, the news from this "small user" are started to be 
spread by many other users, Fig.~\ref{interaction1}.
\\
Nowadays thanks to the advantage of telecommunications, 
huge amount of  data opens great opportunity 
to understand processes in society and estimate models of them,
which was not possible 
before. 
But even before the accessibility of such data 
it was possible analytically to estimate cognitive properties of society. 
For instance, the sociologists 
P.Killworth and R.Dunbar defined and estimated a limit to the number 
of people with whom one can maintain stable social relationships, the so-called, Dunbar's number \cite{Dunbar1992}. 
Development of social models including accessible applications to 
studies of \emph{ "flows"} of opinion in society,
riot behavior, innovation, strikes, 
voting and  migration, 
have been extensively deliberated during the last decades 
\cite{Granoventer, Gonzalez-Bailon2013, heitzig2012}. 
From the series of seminal works it became clear that 
opinion "flows" 
are mainly governed by the "network hubs", 
yet in \cite{Kitsak2010} it was found that node degree 
is not the only characteristic of the "node importance". 
Instead, the most efficient spreaders 
are those located within the 
core of the network \cite{seidman1983}. 
There is a number of examples demonstrating that  
information propagation can be represented as \emph{ "flows"} 
or "opinion waves". 
An illustrative example of such flows is the circulation of ideas 
among articles through 
citation network \cite{Kuhn2014, Lermana}. 
\\

Let us assume that each person is represented as a node 
in a network (toy representation of society). 
Before to come to the main research questions of this chapter, 
let us first compare the information spread model 
with a disease contagion model \cite{lentz2012}, since the opinion spread  
could also be understood as a special type of contagion. 
In a recent work \cite{rosenthal2015} the issue on difference between complex and simple contagion models 
has been addressed: 
in simple contagion models (SIR models) the most influential nodes are typically 
nodes with high degree and low clustering, while 
in complex contagion models the most influential nodes are typically 
characterized by low degree and high clustering.
The main aspects, 
which differentiate various types of \emph{contagion spread mechanisms}, can be 
conditionally separated into:
\\ 
\indent \emph{The mechanism of spreading}, which
is determined by properties 
of spreading, 
the stochastic or deterministic character of the information spreading, etc. 
\\
\indent \emph{The mechanism of node state change}, 
which determines how each node changes its state 
with dynamics on the network, including 
for example, resistance to change its current state.
\\
In the series of recent works \cite{Sayama2014, Ghosh2014} 
it has been found that disease spread is more likely to be 
homogeneous among groups and depends mostly on the properties of the nodes, i.e. on \emph{mechanism of node state change}. 
But on the other hand, the speed of opinion circulation strongly 
depends on the social group properties, society structure and \emph{the mechanism of spreading} 
\cite{Zanette2006, Anderson2015, Belik, romasco2015}. 
Ties strength is important for social contagion which can be modeled as  
the status difference of the nodes \cite{Granoventer},  
one of possible examples is a weighted voter model \cite{Kimura2010}. 
This gave motivation to design a particular novel type of DN model, the HOpS model
where a heterogeneity parameter plays an important role in dynamics of the model. 
The HOpS model I define in Subsection \ref{model_set} after formulating types of dynamical network models
and the research questions.

\subsection{Research questions: graph dynamical models}
\begin{figure*}[h]
\includegraphics[width=0.4 \textwidth]{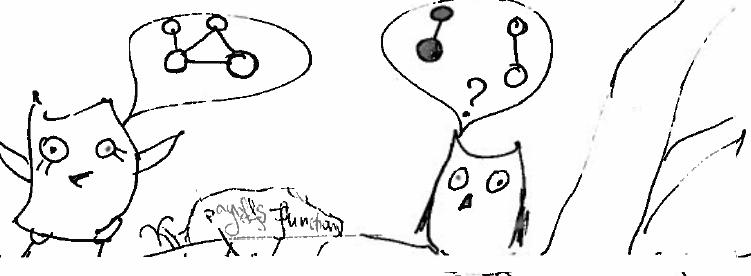} \centering 
\caption{{\bf Illustration of dynamical network model  
with different types of consensus states}: 
total consensus (on the left) or separation into smaller groups (on the right), 
and in each of them consensus is reached.}
\label{adapt_owl}
\end{figure*}

Ultimately, 
in this chapter we shall deal with the \emph{following research questions} to study 
DN models: 
\\
\indent 1. What are the conditions for a dynamical network model to come to a consensus state (defined further)?
What is the speed of convergence towards the consensus? 
Is the consensus state unique, Fig.~\ref{adapt_owl}? 
How to characterise properties of phase space of dynamical network model analytically? 
\\
\indent 2. Is it possible to estimate the model evolution 
on a certain network topology without numerical simulations, for instance, using the transformation operator approach?
\\
\indent 3. 
How is the underlying network topology reflected in the model's dynamics? 
Are there any network topologies for which the model can be completely analysed? 
What are the effects of heterogeneous spread of opinion on the network?
\\ 
All in all, I examine behavior of the HOpS model looking at these questions. 
Further I formulate a brief classification of DN models, 
Section \ref{dyn_netw_mod}, 
and methods overview to the existing methods in Subsection \ref{all_analyt_met}.

\section{Dynamical network models classification} 
\label{dyn_netw_mod}

\begin{figure*}[h]
\includegraphics[width=0.8 \textwidth]{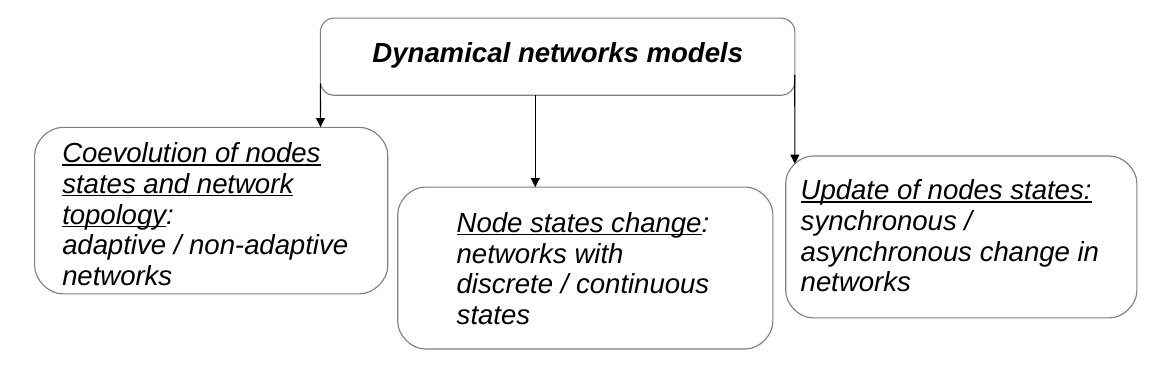} \centering 
\caption{{\bf Properties of dynamical networks models are divided into several classes} 
depending on characteristics of functions, acting on node states, type of coevolution 
between nodes states and network topology.}
\label{scheme}
\end{figure*}

Any decent classification allows to find out, what has been done in a field and what remains
undiscovered, however, recently new DN models on networks have not be structurally classified.  
Social, economic 
and biological networks can be modeled as dynamical 
network models or agent-based models \cite{Boguna2004, Bornholdt2000, Colliza, Holme2006, Botta2011}, 
where 
each node has the possibility to change its state. 
Looking into real-world complex networks 
one can find many instances of networks whose states
and topologies coevolve, i.e. they interact with each other 
and keep changing 
often over the same time scales, due to the system dynamics
\cite{Sayama2013, heitzig2016}. 
Depending on a type of a process being modeled, 
nodes of the dynamical network can have \emph{discrete and/or continuous} evolving states, 
\emph{adaptive and/or non-adaptive} strategies of evolution, and 
simultaneous switch of states of all nodes or a switch in a randomly picked node. 
Following such an approach 
divides properties of DN models of into classes, shown in Fig.~\ref{scheme}. 

\subsubsection{Graph theoretical notations for dynamical network models}
Let us describe the classification of DN models   
using graph theoretical notations, as it has been done in Chapter II for evolving networks. 
Let us denote a dynamical network model on an underlying graph 
$G=G(V,E)$  as $G(V, C(t), E)$, or simply as $G(t)$, where sets $V$ and $E$ are sets of nodes and 
edges correspondingly, number of nodes $|V| =N$, 
$C(t)$ is a set of nodes' states at time step $t$, where a state of node $i$ 
is denoted by $c_i(t), i\in[1,N] $, where $c_i(t)$ takes values from a fixed set $Q$. 
For simplicity we fix sets $V$, $E$ and consider only finite subsets of integer numbers, which 
can have possible values. 
Let us denote a function $F$, acting on a set of nodes' states. 
In fact, this can be also written in matrix notations. 
Let us denote $C(t)$ as a vector state of enumerated nodes's states at time $t$ 
and $F$ is a matrix, defining transformations of nodes' states 
(the exact form of this is given in Subsection \ref{anal_sol}). 
Then the evolution of dynamical network 
can be written according to a formula $F(C(t)) = C(t+1)$, 
or in other words the evolution of a whole DN model on a static network 
topology can be written as: 
\begin{eqnarray}
F(G(V,C(t),E))=G(V,F(C(t)),E) =G(V,C(t+1),E).
\label{netw_ev}
\end{eqnarray}
Generally, each node of a DN model $G(t)$ 
may have several types of characteristics, instead of only one type $c_i(t)$. 
This can be encoded using additional 
set of nodes' states $\{C^1,... C^k\}$, each set $C^j$ 
for $j^{th}$ type of nodes' characteristics, 
so that then a DN model would be denoted as $G(t)= G(V,C^{1}(t),...C^k(t),E)$. 
\\
As an \emph{example}, 
let us consider an evolving DN model $G(t) = G(V,C(t),E)$ with fixed set of nodes, set of edges 
and boolean set of nodes' states $Q=\{0,1\}$. 
Let a deterministic rule of $G(t)$ be: 
at each time step $t$ a state of each node $c_i(t), i\in[1,N]$ 
is changing its value to opposite value: $0 \rightarrow 1, 1 \rightarrow 0 $. 
If a node is changing its state to an opposite one a function $F$ can be written 
as $F(c_i(t+1)) = (c_i(t)+1)mod 2$. 
\\
Let us consider properties of functions, acting on set of nodes' states.
Let us assume that function $F_i$ acts on some subnetwork $G_i \subseteq G$. 
Assume, 
that a function $F$ is a composition of functions: $F = F_1  F_2... F_n$.  
Notably, function $F_i$ does not necessarily commutate with function $F_j$, 
acting on another subnetwork $G_j \subseteq G$.  
Therefore, it is essential to distinguish the order in which functions are applied,  
$F_i  F_j (G)$ or $F_j  F_i (G)$. 
\\
Hence, characteristics of function $F$, 
such as in Eq.~(\ref{netw_ev}), typify the evolution of DN models. 
In particular,  
$F$ may act on a set of nodes' states depending on edges evolution or independently on
edges evolution (adaptive/non-adaptive networks); 
values of $C(t)$ may evolve in discrete or continuous time; 
$F$ may act on the whole network, or be applied to separate subgraphs of a network 
using synchronous or asynchronous update mechanism. 
\\
The model is evolving until a DN model reaches a final configuration, 
which 
can be either a consensus for the whole network or consensus, reached 
in disconnected small subnetworks, Fig.~\ref{adapt_owl}. 
Furthermore, in Subsection \ref{anal_sol}  
I look at problems of DN models from another perspective of 
so-called sequential dynamical systems (SDSs) \cite{Barrett2000, Macauley2000}, 
which helps to describe a discrete phase space of DN models. 


\subsection{Techniques to describe dynamical networks models 
}
\label{all_analyt_met} 

There exist various methods to study DN models, 
such as {transfer operators approach, numerical approach to run models on }
ensembles of random 
topological graphs. 
Various random network models  
provide an efficient laboratory for testing various 
collective phenomena in statistical physics of 
complex systems, and are, on the other hand, tightly linked to
statistical, topological properties of random  matrices, 
for instance, vertex degree distribution, 
clustering coefficients, "small world" structure and spectra of adjacency matrices 
\cite{Avetisov2015, Krapivsky2008, Baxter2009}. 
\\
In order to illustrate one quite famous approach 
for studying dynamical system with discrete phase space 
let us consider 
the discrete transfer operators approach to study  Moran model \cite{Moran58}. 
At each time step a random individual of one of two types $ A$ or $B$ 
is chosen for reproduction and a random individual 
is chosen for death; thus ensuring that the population size remains the same 
(the number of individuals is conserved). 
Here Moran process  
can be used to analyze variety-increasing processes such as mutation, as well as 
variety-reducing effects such as natural selection. 
This process describes the probabilistic dynamics in a population of finite constant size 
when two species $A$ and $B$ are competing for dominance.  
Now let us come to the transition matrix method to describe such systems.\\
The main idea of the \emph{transition matrix method} 
is as follows. 
Let us consider system with discrete number of states, numerated by finite numbers $0s< i < N+1$. 
A 
phase space (or state space) is one-dimensional and
discrete set $\{1,...,N\}$, where $N$ is the number of possible states
(in other words, the cardinality of the phase space). 
The so-called transfer operator is thus represented by an $N\times N$ transition
matrix $P$ acting on a vector in $\{0,1\}^N.$
Each entry $P_{i,j}$ of a \emph{transition matrix} $P$ denotes the probability to go from state $i$ to state $j$. 
To understand the formulas for the transition probabilities one has to look at 
the definition of the 
process which states that always one individual will be chosen for reproduction 
and one is chosen for death, i.e. $P_{i,j} \in [ 0,1]$. 
Once all $A$ individuals have died out, they will never be reintroduced into 
 the population since the process 
 does not model mutations 
 and thus $P_{1,1}=1$. 
 For the same reason the population of $A$ individuals will always stay 
 $N$ once they have reached that number and taken over the population and thus 
 $P_{N,N}=1$. Then states $1$ and $N$ are called \emph{absorbing} while the states 
 $2, ..., N - 1$ are called \emph{transient}.  
Analysing properties of transformation matrices, one can study 
possible states of the system. 
\\
Further I introduce the method based on analysis of transformation matrices,  
illustrating it on the Heterogeneous Opinion Status model (HOpS), Subsection \ref{model_set}. 

\subsection{Heterogeneous Opinion Status (HOpS) model setup} 
\label{model_set}
\begin{figure*}[h]
\includegraphics[width=0.65 \textwidth]{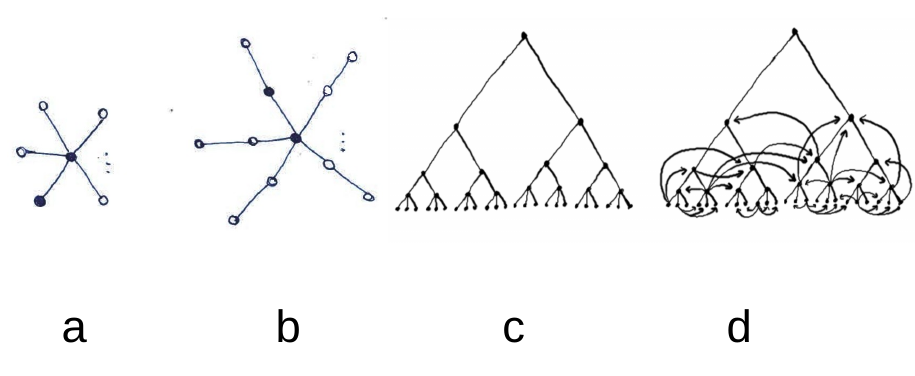} \centering 
\caption{{\bf Star-like networks}: simple star $k =5$ (a), complex star (b). 
Hierarchical networks of different types: 
symmetric tree (c) tree with additional links between different hierarchal layers (d).}
\label{hierar}
\end{figure*}

A novel dynamical network model,
\textbf{H}eterogeneous \textbf{Op}inion \textbf{S}tatus (HOpS) model  
has the following properties, which
allows  to demonstrate analytical methods to characterise 
dynamics on networks. 
\begin{figure*}[h]
\includegraphics[width=0.5 \textwidth]{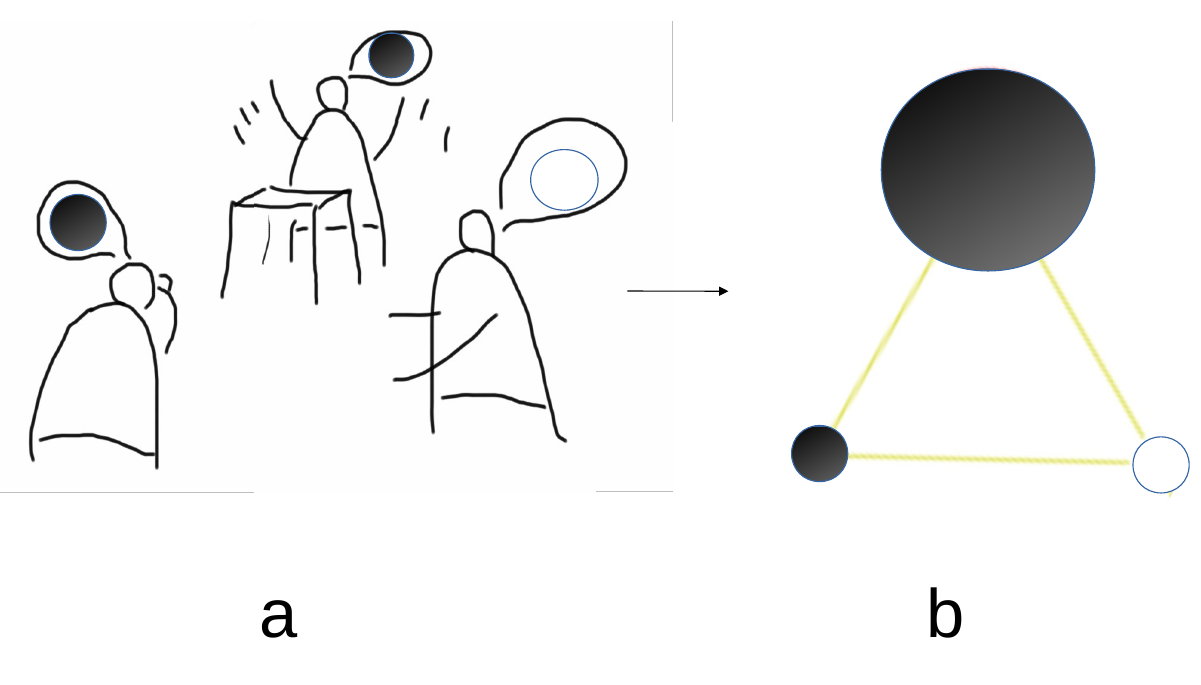}\centering
\caption{ {\bf Representation of the group of people with two possible opinions} marked in 
black and white colors (a). 
Heterogeneous Opinion Status (HOpS) model representation 
of the group from three people (b), 
where the node status is represented by the node size.}
\label{opin_choice}
\end{figure*}
\\
\paragraph{The HOpS model setup}. 
Let us consider the network where each node $i$ has two variables: status and opinion.
Status is fixed and is denoted by a finite number. 
Each node $i$ at time step $t$ has opinion $op_i(t) \in \{0, 1\}$, 
$0$ is encoded as white color of the node and $1$ is black, Fig.~\ref{opin_choice}, 
and $op_i(t)$ changes according to a stochastic rule 
representing the imitation of opinion. 
\\
\textbf{\emph{Definition}}. 
Node $i$ is called an \emph{active node} in the HOpS model, 
if node $i$ is randomly chosen at time-step $t$ 
with its random neighbor $j$ and then 
active node $i$ is changing its opinion $op_i(t)$ 
to opinion of its neighbor $op_j(t)$ 
with a fixed probability, dependent on the difference in statuses. 
\\
\textbf{The algorithm} of  the \emph{HOpS model time step} is given in Table \ref{tabl_mod_time_st}. 
The dynamics is stochastic: 
at each time step a random individual $i$ chooses at random one of the neighbors, node $j$, and accepts the opinion of that neighbor with the probability $p = 0.5\tanh(\sigma(st_j - st_i))+0.5$. 
\\
\begin{table}
\caption {\textbf{Algorithm for the time step of the HOpS model}} \label{tabl_mod_time_st}
\begin{center}
\begin{tabular}{|l|p{10cm}|}
  \hline
  \textbf{Phase of time step} & \textbf{Characteristics of each phase}  \\ \hline 
 1. & Randomly choose an \emph{active node}  $i$ from the network, Fig.~\ref{nonhierar}.
 \\ \hline 
 2. & Randomly choose \emph{one neighbor} of the active node $i$ - node $j$.  \\ \cline{2-2} \hline 
 3. & Change the \emph{opinion of an active node $i$} to the opinion of \\
   & node $j$ with probability $p = 0.5\tanh (\sigma(st_j - st_i))+0.5$. 
\\ \cline{2-2} \hline
 4. & Go to 1., iterating the whole time-step \\
   & until a consensus state is reached.
\\ \cline{2-2} \hline 
\end{tabular}
\end{center}  
\end{table}
\begin{figure*}[h]
\includegraphics[width=0.5 \textwidth]{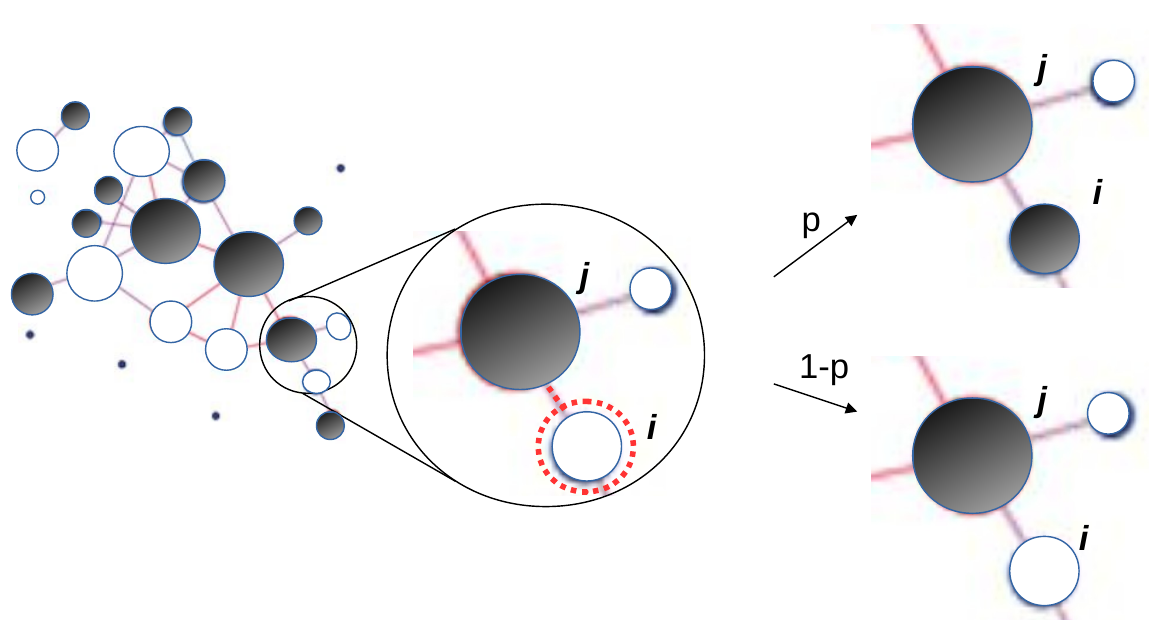}\centering 
\caption{ 
{\bf Illustration of the time-step of the HOpS model}: 
firstly, active node $i$ and its neighbor $j$ are randomly chosen; 
secondly, a state of an active node, opinion $op_i(t)$, is changed with probability  
$p$ depending on the status difference 
for nodes $i$ and $j$. A node status is encoded by a node size.} 
\label{nonhierar}
\end{figure*}
\\
The opinion of an individual can evolve whenever some of the neighbors have opposite opinion. 
Consensus state, when all nodes have the same opinion of two equivalent opinion, either $0$ or $1$,  
is necessarily reached and is the absorbing state of this stochastic dynamics. 
Let us now formulate the HOpS model 
in notations from Subsection  \ref{dyn_netw_mod}. 
The HOpS model is defined on a fixed graph $G(V,E)$  
with a changing opinions of nodes, denoted by  $Op(t) = (\{op_i(t)\}_{i\in[1,N]})$ 
and nodes statuses $St = (\{st_i\}_{i\in[1,N]})$. 
In other words, the HOpS model can be written as $G(V,Op(t), St,E)$. 
\\
\emph{\textbf{The HOpS model input control parameters}} are: 
\\
1) a fixed set of \emph{nodes statuses} $St$; 
\\
2) a distribution of \emph{initial nodes opinions} at time step $t=0$ $Op(0)$, 
(this is discussed in details in Subsection \ref{anal_sol}, and moreover, this, on the first place,  
means that the system is non-ergodic); 
\\
3) a fixed underlying network \emph{topology} $G(V,E)$. 
Moreover, a parameter $\sigma$ of a time step influences the HOpS dynamics, 
its role is discussed separately. 
\\
Note that a probability \emph{function of an opinion change}  
was chosen to be a {sigmoid function $0.5\tanh(\sigma(st_j - st_i))+0.5$}, 
since a sigmoid function represents the increasing likelihood 
of imitation processes to take place with an increase in the status difference \cite{Traulsen2006, endris, Wiedermann2015}. 
By using a status difference, which can be either negative or positive, 
inside $\tanh$-function, we allow asymmetric relations between connected nodes: 
a node with a big status is influenced by the small node less than a node with a small status by a big node. 
\\
It is important to explain a meaning of a novel \emph{heterogeneity} parameter 
of the HOpS model. 
The reason for introducing a new system's "heterogeneity" parameter 
can be seen from the following observation. 
Let us consider
a group with one strong leader-dictator with a very high status,  
where the information transmission 
is directed from the group leader to others, 
in contrary to a homogeneous group. 
By the same token, it has been noticed in \cite{piet1999}   
that a hierarchy in society induces information spread from the leader to others 
more efficiently,  
than in structures where the hierarchical structure is less "pronounced". 
Indeed, a tree-like hierarchical structure  without loops Fig.~\ref{hierar} (a, b, c) 
admits 
smaller speed of convergence towards the consensus state 
than a network with loops, Fig.~\ref{hierar} (d), which is also linked to so-called geometrical frustration \cite{Moesner2006}. 
Note that in the HOpS model it is also assumed that when all statuses are the same, 
an active node changes its opinion to an opinion of its random neighboring node
with probability $0.5$, and with equal  
probability opinion of an active node stays the same. 
\\
To summarize, the HOpS model is a particular kind of DN model with 
two prominent characteristics: 
\\
\indent 
1. \emph{Each node $i$ has status $st_i$}, which is as a characteristic of a so-called social influence. 
A distribution of nodes' statuses introduces heterogeneity to the structure of a DN model 
and to a \emph{mechanism of a node state change}.
\\
\indent 
2. \emph{Opinion change of each node $i$} 
is introduced by a threshold function and induces 
heterogeneity to a \emph{mechanism of an opinion spreading}. 
\\  
In Section \ref{res_hops} I present a new \emph{methodological framework} for a class of dynamical network models. 
This methodics reveals analytic solutions for the HOpS model on symmetric networks. 
As the next step, 
I consider the HOpS model dynamics on random Erd\H{o}s-Renyi networks \cite{Erdos1959}, 
Subsection \ref{num_res}. 

\section{Results for Heterogeneous Opinion Status (HOpS) model}
\label{res_hops}

The \emph{analytical} solutions for the Heterogeneous Opinion Status model 
for particular networks topologies are introduced in 
Subsection \ref{hops_symm}.  
The \emph{numerical} results 
for the HOpS dynamics are described in 
Subsection \ref{num_res}. 

\subsection{Analytic results for the HOpS model}
\label{anal_sol} 
There has been a variety of numerical studies on dynamics on networks, 
while analytic approaches to DN models analysis always have been lacking. 
Here I introduce a novel approach 
to study DN models using notations from theory of generalized cellular 
automata, Markov chains, and illustrating this approach on the HOpS model. 
\emph{The main idea} of this technique is
that for some model configurations, it is possible 
to calculate analytic solutions due to topological properties of these configurations. 
Let us call such configurations \emph{basic configurations}. 
Which are these configurations? 
It is natural, 
first to consider   
basic network structures, particularly, 
a class of \emph{symmetric networks}.  Then further one can generalize 
model solutions for more complex underlying networks. 
The intuitive notion of 
a \emph{graph symmetry} can be detected by graph measures \cite{Holme2006b} 
and is characterized by features of group of graph automorphisms \cite{Harary1969}.  
For symmetric networks this group is non-trivial \cite{Guichard}.   
A formal definition for symmetric graphs is as follows (here the property of symmetry is defined for $G(V,E)$). 
\\
\\
\textbf{\emph{Definition}}. 
Two nodes $u$ and $v$ of a graph $G$ are \emph{similar}, if for some automorphism $\alpha$ 
of $G$, $\alpha(u) = v$. A fixed point is not similar to any other point. 
Two lines $x_1 = u_1v_1$ and $x_2 = u_2v_2$ 
are called similar if there is an automorphism $\alpha$ of $G$ 
such that $\alpha(\{u_1, v_1\}) = \{u_2, u_2\}$. 
Only graphs without isolated points are considered. 
A graph is \emph{point-symmetric}, if every pair of points are similar; 
it is \emph{line-symmetric} if every pair of lines are similar; 
and it is \emph{symmetric} if it is both point-symmetric and line-symmetric \cite{Harary1969}.  
\\
\\
Coming back to notations in Section \ref{dyn_netw_mod}, 
a \emph{state of the HOpS model} at time step $t$ is denoted as $G(V,Op(t),St,E) $ 
and is determined 
by set of nodes' states $Op(t)$.  
The opinion distribution $Op(t) = \{op_i(t), i\in[1,N]\}$ are components of a state vector 
at each time step.   
The state vector $Op(t)$ depends on 
a fixed statuses distribution $St$, a network topology $G(V,E)$, 
the initial 
opinions at $t=0$ time step $\{op_i(0), i\in[1,N]\}$ 
and on the time-step characteristics. 
The function $F$
describes a change of state-vector $F: G(V,Op(t),St,E) \rightarrow G(V,Op(t+1),St,E)$. 
Important to notice that function $F$ is contingent on the network topology.
\\
It has been noticed that evolution of the processes  
on symmetric network topologies without loops has peculiar properties \cite{Kohler1990}. 
At the same time, topological properties of networks, 
such as symmetry,
influence the main parameters, 
quantitatively characterize random walk on networks. These characteristics are, 
for instance,  hitting time, 
cover time, mixing rate \cite{Asz1993}.
The classical theory of random walks deals with random walks on simple, 
but infinite graphs, like grids, and usually studies their qualitative behavior: 
does the random walk return 
to its starting point with probability one or if it returns infinitely often? 
Or how \emph{structural or topological properties of networks} 
are related to properties of transformation matrices of random walks
\cite{Dorogovtsev2003ab, Boccaletti2006}? 
An example of random walk properties is the mean quadratic derivation \cite{Klafter2008, thiel2012}, 
the characteristic time, 
i.e. time after which
the random walk has passed through all the nodes, defined for finite networks 
\cite{Bonaventura2013}. 
\\
With this in mind, first, I consider %
the HOpS model dynamics on symmetric networks without loops, for which 
I use 
the random walk theory \cite{nechaev2003, Noh2004, sokolov2012} 
and demonstrate the HOpS model results, conducted using picture of discrete-time 
random walk, Subsections \ref{hops_symm} and \ref{hops_star}.

\subsection{The HOpS model dynamics on linear networks}%
\label{hops_symm}
As a starting point, 
I reveal analytic solutions for the HOpS model for particular kinds 
of symmetric networks: 
linear and star-like networks. 

\subsubsection{Analytic solution for the HOpS model on linear networks}
Here I consider the HOpS model on linear networks, explained in two following propositions. 
Further term "model" is meant to be the HOpS model if not stated otherwise. 
\\

\begin{figure*}[h]
\includegraphics[width=0.3 \textwidth]{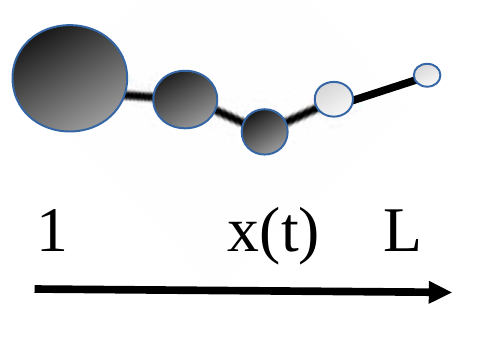} \centering 
\caption{{\bf The HOpS model on a linear network} for $L=5$ nodes. 
A position of a border $x(t)$ between black and white nodes 
($0< x(t)<L+1$) is considered as a random walker.  
The special initial condition: node statuses decrease linearly, 
nodes with different opinions are separated by a border. 
The HOpS model, starting from such initial condition, 
reaches one of two final stable states: all nodes have the same color.}  
\label{chain}
\end{figure*}

\begin{figure*}[h]
\includegraphics[width=0.71 \textwidth]{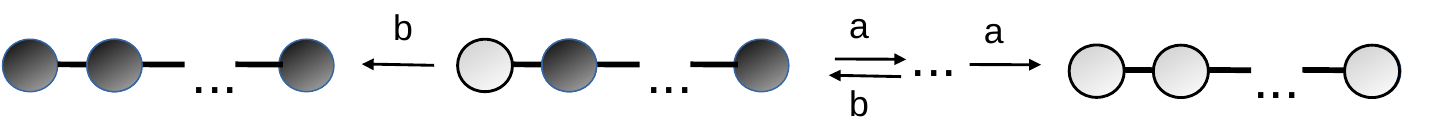}\centering
\caption{{\bf Schematic diagram characterizes a phase space 
of the HOpS model on a linear network for a special initial condition. } 
The phase space is presented as a sequence of states and transitions between them, shown by arrows.
Here the phase space is shown only for the HOpS model 
with a special initial state: $L-x(0)$ nodes from the left border are black,
and $x(0)$ nodes from the right border are white at time step $t=0$. 
}
\label{diag_lin_net}
\end{figure*}

Let us consider the HOpS model on a \emph{linear network} of length $L$. 
A space of all possible states of the HOpS model is denoted by $\mathbf{S}$, where each model 
state is fully described by 
opinion state vector is $Op(t) = (op_1(t),...,op_N(t))$, where 
$op_i(t)$ is opinion of a node $i$. 
Starting from random initial conditions 
a phase space $\mathbf{S}$ has $2^L$ possible distinguishable states. 
\\
\textbf{\emph{Definition}.} \emph{A state of the DN model} 
at time step $t$ is a state vector of opinions 
$Op(t) = (op_1(t),...op_N(t))$, when other model characteristics  are fixed,  
such as status distribution $St = (st_1,...st_N)$ or network topology.
\\
Let us now assume, that the model starts from 
a \emph{special initial model configuration}: 
$Op(0)= (op_1(0), ... op_N(0)) = (0,0,...,0,1,...1)$ such that 
$x(t)$ left nodes are black, 
$L-x(t)$ right nodes are white. 
Moreover we assume, that nodes' statuses are linearly decreasing from black to white nodes, Fig.~\ref{chain}, 
such that $\forall i \in [1,L-1]$ 
a status difference is fixed  $(st_i-st_{i+1}) = \Delta_{st}$.   
\\
\\
\textbf{Proposition I.}
\\
Starting from a \emph{special initial model configuration},
all model states  belong  
to a subspace $S'$
of a phase space $\mathbf{S}$: $S' \subset \mathbf{S}.$ 
Such a subspace is called an invariant subspace, since  
it fulfills the condition, that for 
any vector-state $Op(t) \in S':$ $Op(t+1) \in S'$ $ \forall t$. 
The number of states in this subspace $|S'| = L$.
\\
\emph{Proof}: 
The number of states for the invariant subspace $S'$
equals the number of all possible positions of the border $x(t)$
between black and white nodes. 
Starting from the special initial condition, 
the model is able to reach only a subspace of all system states, 
which belong to so-called invariant subspace.  
Hence, finding an invariant subspace of 
the system allows to describe all possible model states, or in other words, 
full phase space. 
\\
\\
\textbf{Proposition  II}. 
\\
The HOpS model dynamics with the special initial condition 
is equivalent 
to dynamics of an asymmetric bounded random walk $x(t)$ on a linear network. 
\\
\emph{Proof:}
Let us consider  
the probability of any black node to be converted into a white node 
is equivalent to $(0.5\tanh\sigma \Delta_{st}+0.5)$, 
where $\Delta_{st} =st_i-st_j$ 
is a fixed status difference. 
Then a state of the whole system is described just by position of a random walker  $x(t)$. 
A probability of a random walk to drift to the right is denoted by 
$a=0.5\tanh(\sigma  \Delta_{st}) +0.5$ and 
probability of a random walk drift to the left is denoted by
$b = 1- a$. 
The model has two consensus states: when all nodes are either all black or all white. 
A probability of a random walker to reach the right border 
is equal to a probability of the HOpS model to come to a consensus when all  
nodes are black. 
\\
\\
Here I refer back to the research questions on DN models behavior, Subsection \ref{res_ques}, 
which are translated to the language of the random walk theory. 
\\
\\
\textbf{Bounded asymmetric random walk on a linear network} 
\\
As it has been previously shown, 
the HOpS model dynamics on a linear network with 
a special initial configuration, as in Fig.~\ref{chain}, 
is described by a \emph{ random walker} $x(t)$. 
The probability of a random walker $x(t)$ to be shifted to the right  
equals $a$, 
and the probability of a random walker $x(t)$ to be shifted to the left equals $b$, as in Proposition II.
Then probability $p(x(t+1) = i)$ 
for an \emph{asymmetric random walk} $x(t)$ 
to be in position $i$ at time step $t+1$ can be written as: 
\begin{eqnarray}
 p(x(t+1)=i) = 
\left \{
  \begin{matrix}
0 : | x(t+1) - x(t)| >1\\
a: x(t+1) - x(t) =1\\
b: x(t+1) - x(t) =-1\\
1-a-b: x(t+1) - x(t) = 0
\end{matrix}
\right \}
\label{rw_matr}
\end{eqnarray}
For convenience let us set $a+b =1$, 
which corresponds to a case when a random walker cannot stay on the same node.   
All together, this defines a transformation matrix $P$ with 
size $|P|=|S'| \times |S'| = L\times L $. 
The non-zero entries of a matrix $P$ are values on diagonals parallel to the main diagonal. 
Then an evolution equation for state vectors is defined 
by a tridiagonal right-stochastic matrix $P$: 
\begin{eqnarray}
Op(t+1) = Op(t) \cdot P
\label{evol_eq}
\end{eqnarray}
where $Op(t)$ is a state vector of opinions at time step $t$, 
and $P$ is a transformation matrix (column-stochastic) of a corresponding Markov chain. 
Hence, estimating asymptotics of the HOpS model 
is equivalent to Gambler's ruin problem \cite{Weber2012},  
which describes  
an asymmetric random walk on the integers $(1, . . . ,L)$, 
with absorption at $1$ and $L$ nodes. 
Solving the Gambler's ruin problem, 
we find solutions for the HOpS model on linear networks, as described below. 
\\
\\
\textbf{Proposition III.}
\\ 
Let us consider a bounded random walker on $[1,L]$ interval, starting from position $x_0$
with probability $a$ to walk to the right and probability 
$b$ to walk to the left. 
Then an asymptotic \emph{solution for an asymmetric bounded random walk on a linear network}
is given by a probability to hit the right border:
\begin{equation}
p(x_0,a) = \frac{(a^{x_0} (1-a)^{L-x_0} - a^L)}{((1-a)^L - a^L)}.
\label{prob_rw}
\end{equation} 
\\
\emph{Proof:} 
Let a random walker be initially in position $x_0$.   
$p_i(j)$ defines a probability starting from $i$ to hit $j$.  
It is easy to see, that $p_0(0) =1, p_0(L) =0$ and 
correspondingly $p_i(0) = a p_{i-1}(0) + bp_{i+1}(0)$.  
Then a characteristic equation 
is 
$$ax^2 -x + b = (x-1)(ax-b)$$ 
which has roots $\{1,b/a\}$. 
For $a=b$ a random walker becomes symmetric. 
For $a\neq b$ a general solution $p(x_0,a)$ is sum of the roots 
with the coefficients defined by the absorbing states at $1$ and $L$. 
Thus 
the probability of a random walker to hit one of the borders is: 
\begin{equation}
p(x_0,a) = \frac{(a^{x_0} (1-a)^{L-x_0} - a^L)}{((1-a)^L - a^L)}.
\label{prob_rw}
\end{equation}
Moreover the Gambler's ruin problem can be viewed as a special case of a first passage time
problem, 
which asks to compute the probability that a Markov chain, initially in state, hits one fixed state before
another. 
\\
\begin{figure*}[h]
\includegraphics[width=0.54 \textwidth]{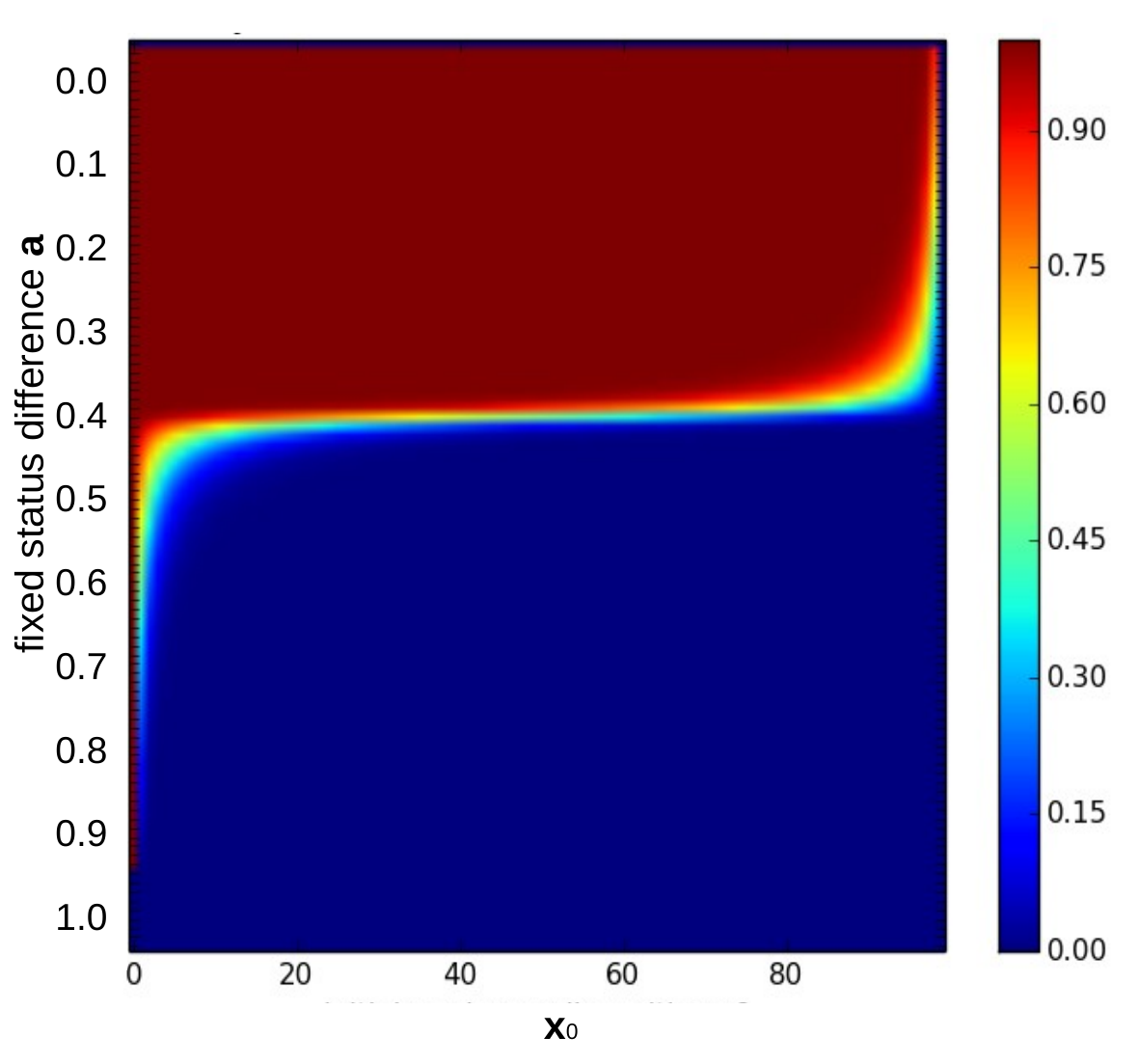}\centering 
\caption{{\bf Dynamics of the HOpS model on a linear network.} 
$x_0 \in [0,100]$ (horisontal axis) is initial number of black nodes 
and $a=0.5\tanh(\sigma \Delta_{st})+0.5$ (vertical axis). 
Colorbar corresponds to the probability $p(x_0,a)$ for the system 
to come to one certain consensus, when all nodes are black, starting from $x_0$. 
}
\label{prob_line}
\end{figure*}
\\
To sum up, the analytical results for the HOpS model on a linear network are:
\\
(i)
\emph{The HOpS model dynamics on a linear network is described by  Propositions I, II, III.}
The quantitative characteristics of a phase space of the model are given in Proposition I. 
The 
formula (\ref{prob_rw}) estimates 
the probability to reach a stable states of the model. 
From the formula (\ref{prob_rw}) it is clear that 
$a =0.5\tanh (\sigma \Delta_{st})+0.5$ characterizes a speed 
of the model convergence towards a consensus state and $\sigma$ 
denotes scaling of a spreading process on a network. 
\\
(ii) 
\emph{The analytic result of Proposition III is illustrated by 
the numerical result, Fig.~\ref{prob_line}. }
Each model simulation is made for values of $x_0$ and $a$. 
$x_0\in [0,100]$ corresponds to initial number of black nodes. 
$a =0.5\tanh(\sigma \Delta_{st})+0.5 \in [0,1]$ characterizes the status difference between nodes. 
Then a probability to find the model in one certain stable state 
numerically corresponds to a ratio between a number of model simulations, 
which reach one certain possible consensus state 
to a number of total model simulations. 
The probability to find the model in its final state 
is marked by the color of each point $(x_0, a)$, Fig.~\ref{prob_line}. 
Red region above the yellow curve on Fig.~\ref{prob_line} 
corresponds to model simulations when the model converges towards a consensus,  
or in other words, a random walker reaches the right border. 
The curve separating red and blue regions 
is implicitly defined via relation $p(x_0,a)=0.5$, Eq.(\ref{prob_rw}), 
which gives the formula for the curve: 
$a^{x_0} (1-a)^{L-x_0} =2(1-a)^L - a^L$. 
Blue region below the curve
corresponds to another absorbing state 
when the model reaches another stable state and, hence,  a random walker
with the characteristics from that region 
never reaches the right border. 
\\
(iii) 
\emph{ The schematic diagram of a discrete phase space of the HOpS model on linear networks   
is presented in Fig.~\ref{diag_lin_net}. }
The arrows on the diagram correspond to transitions between different model states. 
Topology of a diagram of the model on a linear underlying network is trivial, yet  
it illustrates how one can represent a part of a phase space of DN models. %
For more convoluted underlying network topologies 
the model phase space has more complex structure, as it is shown in Subsection \ref{hops_star}. 
\\

\begin{figure*}[h]
\includegraphics[width=0.58 \textwidth]{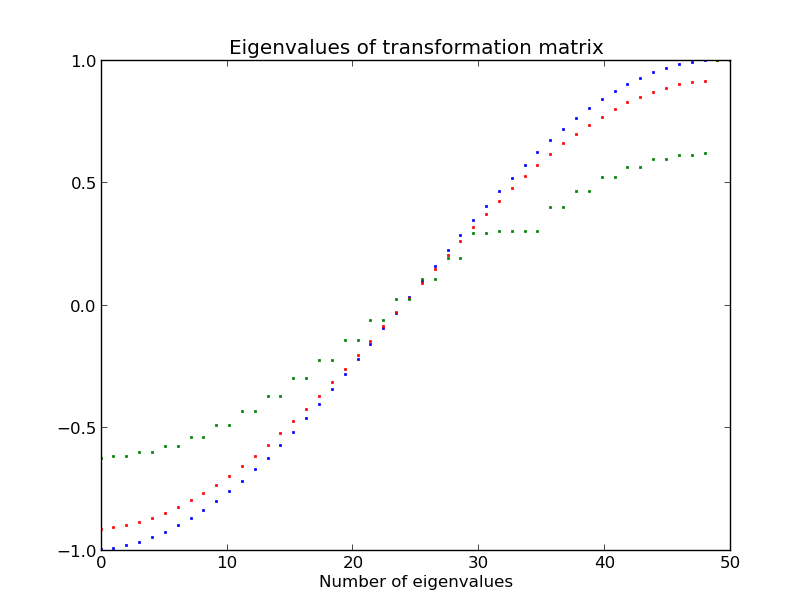}\centering
\caption{{\bf Spectrum of a transformation matrix for the HOpS model on a linear network}
for $L=50$. 
Larger spectral gaps are observed for larger 
parameter values $a = 0.5 \tanh(\sigma \Delta_{st}) +0.5$ values,  
as the result, this gives faster convergence towards the consensus. 
Spectra for $a=0.5$ is blue dashed line, for $a=0.7$ - red dashed line,
for $a = 0.9$ - green dashed line. } 
\label{spectrum}
\end{figure*}

Additionally to the analytical results, 
spectra of transformation matrices $P$ 
for various values of parameter $a = 0.5\tanh(\sigma \Delta_{st})+0.5$ 
are calculated in Fig.~\ref{spectrum}. 
Interestingly,
spectral properties of a transformation matrix and 
mixing properties of the system, described by this transformation matrix, are related. 
The \emph{spectral gap}, 
by definition, is a gap between the
largest and the second largest eigenvalues of a matrix. 
As can be observed from Fig.~\ref{spectrum},
the spectral gap is smaller for larger $a$ values ($a>0.5$), which 
means that larger  values of parameter $a$  
correspond faster mixing times of the system \cite{Chekroun2014} 
and forces faster reaching the consensus than for smaller $a$ values.  
Translating this to the language of the HOpS model, 
the bigger the status difference $\Delta_{st}$, 
the faster the equilibrium state is reached. 
This property is also related to the mixing time of the corresponding Markov chain 
and it  is also known as Cheeger Inequality \cite{Asz1993}. 
\\

\subsection{The HOpS model on star-like networks} 
\label{hops_star} 

After demonstrating analytical solutions for the HOpS model dynamics 
on a linear network, 
the next step is to consider the HOpS model on more general 
symmetric structures, such as star-like networks,  Fig.~\ref{hierar} (a,b). 
First I consider particular types of  \emph{star-like networks}. 
\\
\textbf{\emph{Definition.}} 
A \emph{simple star} is a network 
with one central node and 
$k$ "leaves" i.e. one-node edges, attached to a central node, Fig.~\ref{hierar} (a). 
A simple star is a tree-like network 
with tree depth $1$.
\\
It is 
important to emphasize, that 
the definition of a star-like network 
highlights two main differences in comparison with 
a linear network:  
(1) We have to 
cope with a more complex network topology. Any type of dynamics on the star-like graph 
is obviously not equivalent to dynamics in the case of 
linear network \cite{Kromer2015}. 
(2) As the consequence, the over-all complexity of the model dynamics 
on star-like network is larger.
However, as it is found below, the analytic techniques to describe 
the model dynamics on simple star-like networks 
are originated from the framework  
for the linear network case, Subsection \ref{hops_symm}. 

\begin{figure*}[h]
\includegraphics[width=0.3 \textwidth]{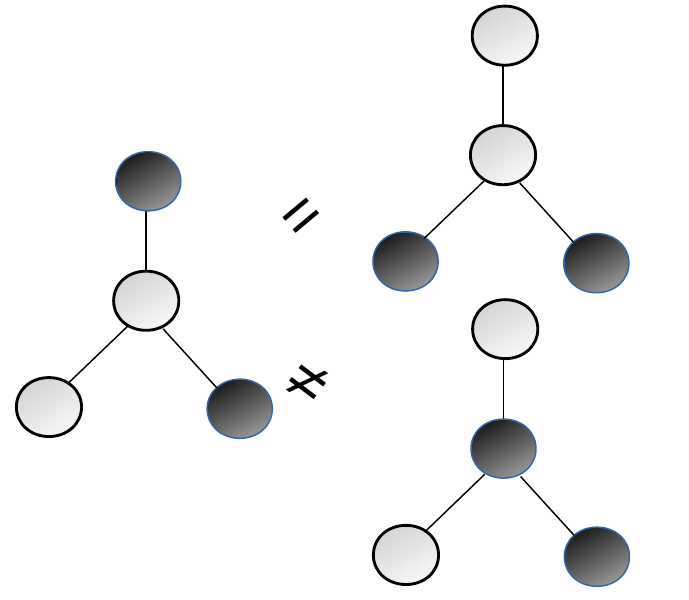}\centering
\caption{{\bf Equal states of the HOpS model on a \emph{simple star}} with $k = 3$ 
leaves: two states are equivalent iff in both states the central node 
has the same color and  
the number of nodes-leaves with identical color is the same.}
\label{sim_star}
\end{figure*}

\subsubsection{The model dynamics on a simple star network}
Let us first consider a \emph{simple star network} 
with $k$ one-node "leaves" and one central node. 
The number of the system's states for random initial conditions is  
$|\mathbf{S}| = 2^{k+1}.$ 
Now assume, that all "leaves" of a simple star network have some fixed statuses 
$st_i = s$, $i=\{1,...k\}$, 
and a central node has a higher status $st_{k+1} = s+\Delta s$ 
where $\Delta s$ is a parameter of status difference. 
As we saw, the HOpS model on a linear network, 
the existence of the invariant subspace 
simplifies the description of the whole discrete phase space $\mathbf{S}$ of the system. 
Remarkably, 
there is no non-trivial invariant subspace inside space $\mathbf{S}$, 
for simple {star-like networks}. 
If there existed such an invariant subspace $S'$, 
then there would be a special initial condition, 
i.e. a vector-state $Op(t)$ from a subspace 
$Op(t) \in {S}'$ such that for  $ \forall t$ a transformation $P$ results in: 
$Op(t)P \in {S}'$. But using finite enumeration method \cite{Harary1967}
it is easy to see that there is no such initial conditions,
in contrary to the case for the HOpS model on a linear network.  
Simply speaking, the reason for this is that 
a structure of a group of symmetries for a star-like network is more complex than a group of symmetries for a linear network. 
Nevertheless, it is possible to "simplify" 
a space $\mathbf{S}$ using a natural algebraic technique 
to induce the parametrization on a space 
$\mathbf{S}$ \cite{vinberg2001}.
The main advantage of the parametrization is that it allows
to change the structure of the space $\mathbf{S}$, so that 
the parametrized space ${S^*}$  
has an invariant subspace, while the corresponding "initial" space $\mathbf{S}$ 
doesn't.
Note also that 
nodes-leaves of a star network do not interact with each other. 
This means that the probability of transition between a state with $n$ black leaves and a state with $(n-1)$ black leaves is independent from $n$.
Let us first consider elements of the space  $\mathbf{S}$, 
the states of the model at time steps $t_i$ and $t_j$ with corresponding 
state vectors  $Op(t_i)$ and $Op(t_j)$. 
In order to parameterize a full space of states $\mathbf{S}$ 
I introduce a natural equivalence relation between states. 
\\
\\
\emph{\textbf{Definition.}} 
I call two states of the HOpS model on the star-like network \emph{equivalent}, 
as in Fig.~\ref{sim_star}, $Op(t_i)\sim Op(t_j)$
iff:\\
\indent 
in both states the central node has the same color;\\
\indent 
in both states the number of nodes-leaves with white color is the same.
\\
\\
Using such an equivalence relation, 
we are now ready to parametrize the space of states $\mathbf{S}$.
Later on I come back to this issue, discussing equivalence of DN models.  
Parameterized space $S^*$ 
is then defined as $S^* = \mathbf{S}/ \sim$. 
Notably, a group of equal states of the space $\mathbf{S}$ corresponds 
to a state of the space $S^*$.
Let us call states of space $S^*$ \emph{macro-states}, 
in order to distinguish them from states of "initial" phase space $\mathbf{S}$. 
In the following proposition I describe equivalent macrostates of the HOpS model from the algebraic point of view.
\\
\\
\textbf{Proposition IV.}
\\
All equivalent macro-states of the model on a star-like network form a group $\Pi'$ in respect to 
the operation of a permutation. 
\\
\emph{Proof}. 
Let us consider 
equivalent states $Op(t_i)$ and $Op(t_j)$ from the whole discrete phase space $\mathbf{S}$ 
of the model on a star-network.  
Each node has a Boolean value $ op_i \in \{0,1\}$. 
Then the proposition follows from the fact  
that vector states $Op(t_j)$ and $Op(t_j)$ 
belong the same macro-state in the parametrized space $S^*$. 
In another words, 
two states $(op^1_1(t_i),...op^1_{k+1}(t_i))$ and $(op^2_1(t_i),...op^2_{k+1}(t_i))$ of space $\mathbf{S}$ belong to the same macro-state of space $S^*$, 
iff there exists such a permutation $\pi$: 
\begin{equation}
\exists \pi \in \Pi':
(op^1_1(t_i),...op^1_{k+1}(t_i))= (\pi(op^2_1(t_i),...op^2_{k+1}(t_i))). 
\end{equation} 
Without loss of generality, let us enumerate  
a central node as the  $1^{st}$ node with opinion, denoted by $op_1(t)$. 
Then a permutation on states of nodes 
$\pi \in \Pi' $, which transforms two equal states between each other,
has a property: 
\\
$\pi (op_1(t_i),op_2(t_i),...op_{k+1}(t_i)) = (op_1(t_i),\pi(op_2(t_i)),...,\pi(op_{k+1}(t_i))) $, \\
so that opinion of a central node stays the same. 
In other words, the group $\Pi'$ consists of such permutations for which 
a value of a central node is preserved.  
\\ 
Interestingly, a subgroup $\Pi'$ of 
a group $\Pi$ of all permutations of a set of $(k+1)$ numbers is isomorphic to a subgroup of symmetric group. 
This follows from the Caley theorem \cite{babai_1995}, 
which states that every finite group $\Phi$ is isomorphic 
to a subgroup of a symmetric group $Sym(\Phi)$. This property of a 
group of permutations gives intuition behind structure of permutations. 
\\
Furthermore,  
an opinion of a randomly chosen \emph{active node} $i$ evolves in time as: 
$ op_i(t+1) = (op_i(t)+1)mod 2$. 
\\ 
The discrete phase space of the model on a star-like network
is shown in a schematic way in Fig.~\ref{variant}, 
where each model configuration can be transported to another configurations with 
probabilities $a = 0.5 \tanh(\sigma \Delta_{st}) +0.5$ or $b = 1- a$. 
In fact, the structure of such graphical diagram
is not occasional, and is connected to algebra of processes \cite{Brinksma} and   
sequential dynamical networks \cite{Barrett2000}.

\subsubsection{Sequential dynamical systems}
\textbf{\emph{Definition.}}
\emph{Sequential dynamical systems (SDSs)} 
are constructed from the following components: 
(1) A finite underlying graph $G$ with vertex set $V = \{1,2, ... , N\}$. 
Depending on the context the graph can be directed or undirected. 
(2) A state $x_w$ for each vertex $i$ of $G$ taken from a finite set of values $K$. 
The system state is the $N$-tuple $x = (x_1, x_2, ... , x_N)$, and $x[i]$ 
is the tuple consisting of the states associated to the vertices 
in the 1-neighborhood of $i$ in $G$ (in some fixed order).
(3) A vertex function $f_i$ for each vertex $i$. 
The vertex function maps the state of vertex $i$ at time $t$ to the vertex 
state at time $t + 1$ based on the states associated to the 1-neighborhood of $i$ in $G$. 
\emph{Stochastic Sequential dynamical system (SSDS)} has the same components as SDS
accept that the update rule has stochastic component \cite{Macauley2000}.
\\
Sequential
dynamical systems may be thought of as generalized cellular automata, 
the main difference between SDS and the DN model is that SDS has a deterministic update rule. 
Similarly to the case of the HOpS model
a structure 
of a phase space of SDS is governed by topological properties of 
an underlying graph $G$, 
vertices' states $\{f_i\}_{i\in[1,N]}$, and the so-called update sequence $\omega$ 
defining the transformation of vertices' states. 
Note that  
for deterministic SDSs, each state in its phase space can be transformed only to one state, 
while 
for stochastic SDSs (SSDS) 
each state does not necessarily have only one possible state, where it can be transformed. 
In the definition of DN model, 
a transformation $F$ is defined on a set of all possible states 
$\Gamma =\{G(V,Op(t), St, E), t\in[0,\infty)\}$ so that $F:\Gamma \rightarrow \Gamma $. 
Note, that a transformation matrix of a function $F$ is denoted by $P$.
Another useful notion from theory of SDSs is a \emph{digraph of SDS}, 
a graph, where each link of digraph is associated with 
a transition between model states in discrete phase space of SDS. 
In the following proposition I explain the connection between SDSs and the HOpS model. 
\\
\\
\textbf{Proposition V}.\\
A phase space of the HOpS model is associated with  a digraph of some stochastic SDS.  
\\
\emph{Proof} of this short proposition follows from the definition of a \emph{stochastic SDS}. 
\\
\\
A phase space of stochastic SDS can be understood as 
a weighted underlying graph, where weights denote probabilities of transformations 
between states of the system. 
\\
\\
\textbf{\emph{Definition.}}
The \emph{basin of attraction} of an attractor in a discrete phase space of SDS 
is the set of all states that eventually end up on this attractor, 
including the attractor states themselves. 
The size of the basin of attraction is the number of states belonging to it. 
\\
\\
In case of the HOpS model on a finite networks a basin of attractor 
consists of states, which can be transformed to absorbing states (consensus states). 
The HOpS model on a linear network or on a star graph has two possible absorbing states.
Which of these states will be reached depends on the initial condition of the model. 
\\
\\
\emph{\textbf{Definition}}. 
\emph{The state space} $\Omega$ of the HOpS model (or of any finite DN model) 
is the finite directed graph (digraph),
where an edge exists between states, if they can be transformed from one to another. 
\\
The following proposition describes the HOpS model from the point of view 
of discrete finite dynamical systems \cite{mortveit2001}. 
\\
\\
\textbf{Proposition VI.}
\\
Since the set of states of the HOpS model (in principle, of any finite DN model) 
on a finite underlying network is finite, any directed
path must eventually enter a directed cycle, called a limit cycle.
\\
\emph{Proof:}
Directed paths in full space of the HOpS states $\Gamma$  
correspond to iterations of a function $F$ on the model's states 
or to state, at the beginning of the path. 
Then because there is a finite number of possibilities 
of paths, at some point path of states comes back to the state, 
where it started. 
In other words, 
because we deal with the map between the set of all states 
$\{0,1\}^N \rightarrow \{0,1\}^N $ 
and the sets of possible states are finite, for any given 
initial condition it must have either a unique fixed-point
or a limit cycle, otherwise the set would have to be non-finite. 
\\
\\
\emph{\textbf{Definition}}. 
Two SDSs are called \emph{isomorphic} if there exists a digraph isomorphism
between the phase spaces of these SDSs. These SDSs are stably isomorphic if there exists a
digraph isomorphism between their limit cycle graphs. 
\\
\\
Finally, let us come back to the HOpS dynamics.
We call two DN network models \emph{equivalent in terms of the HOpS model}, 
if their digraphs of phase spaces of these  
models are isomorphic.
Then Propositions I-VI can be extended to describe the HOpS model 
dynamics on a broader class of underlying networks, than just on linear and star-like 
networks, but also on underlying networks, for which the HOpS 
model phase space has non-trivial invariant subspace. 
The digraph of the HOpS model phase space, Fig.~\ref{variant}, 
has a symmetric structure, which, in fact, is connected to symmetric properties of underlying networks. 
In particular, 
the diagram in Fig.~\ref{variant} exposes two absorbing states, 
 transient states and moreover contains a limit cycle. 
Hence, SDSs theory provides a promising tool 
to describe DN models. 
In the following subsection 
I analyze the HOpS model in terms of transformation matrices.

\subsubsection{Transformation matrix approach for the HOpS model on star-like networks}
\begin{figure*}[h]
\includegraphics[width=0.55 \textwidth]{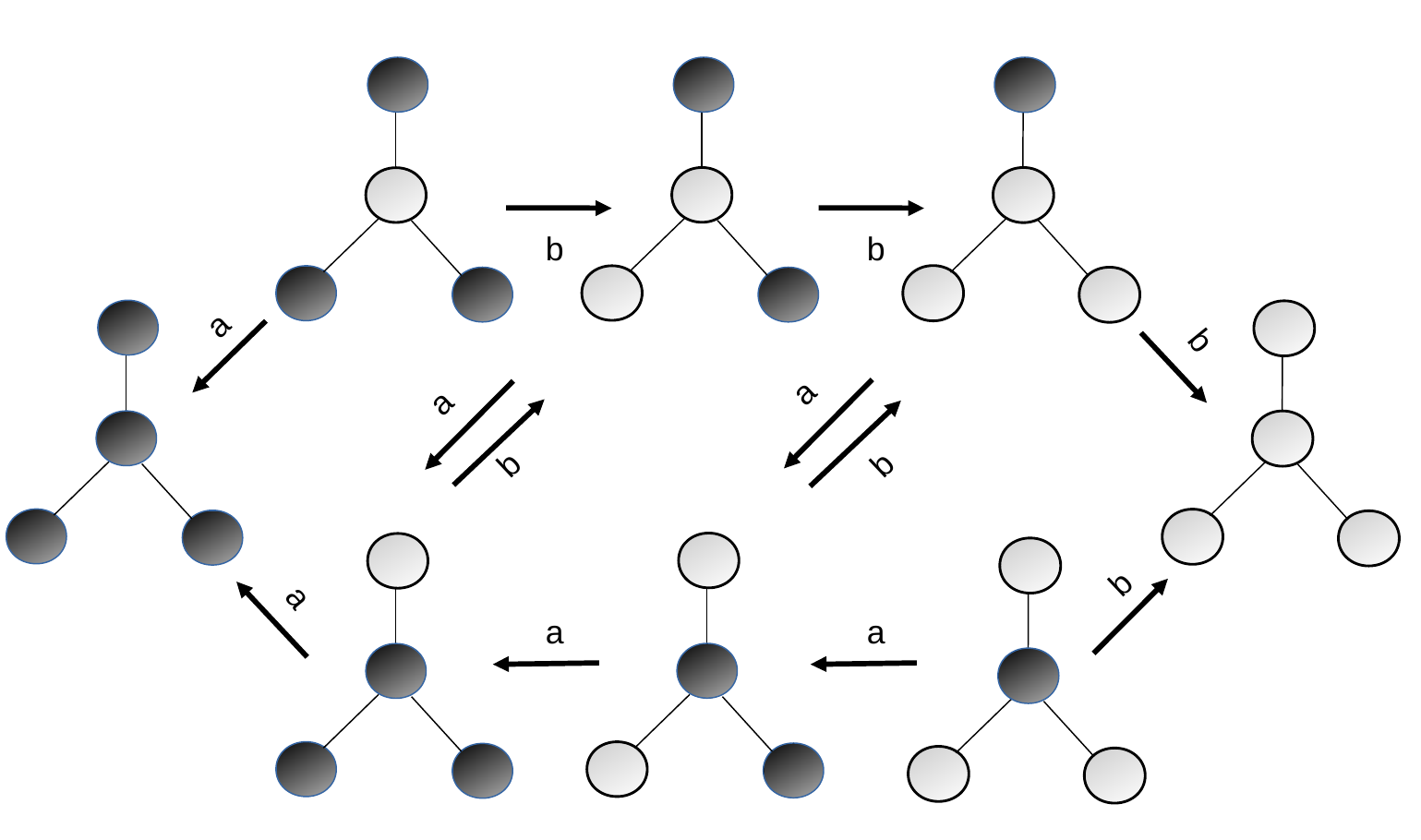}\centering \\
\includegraphics[width=0.55 \textwidth]{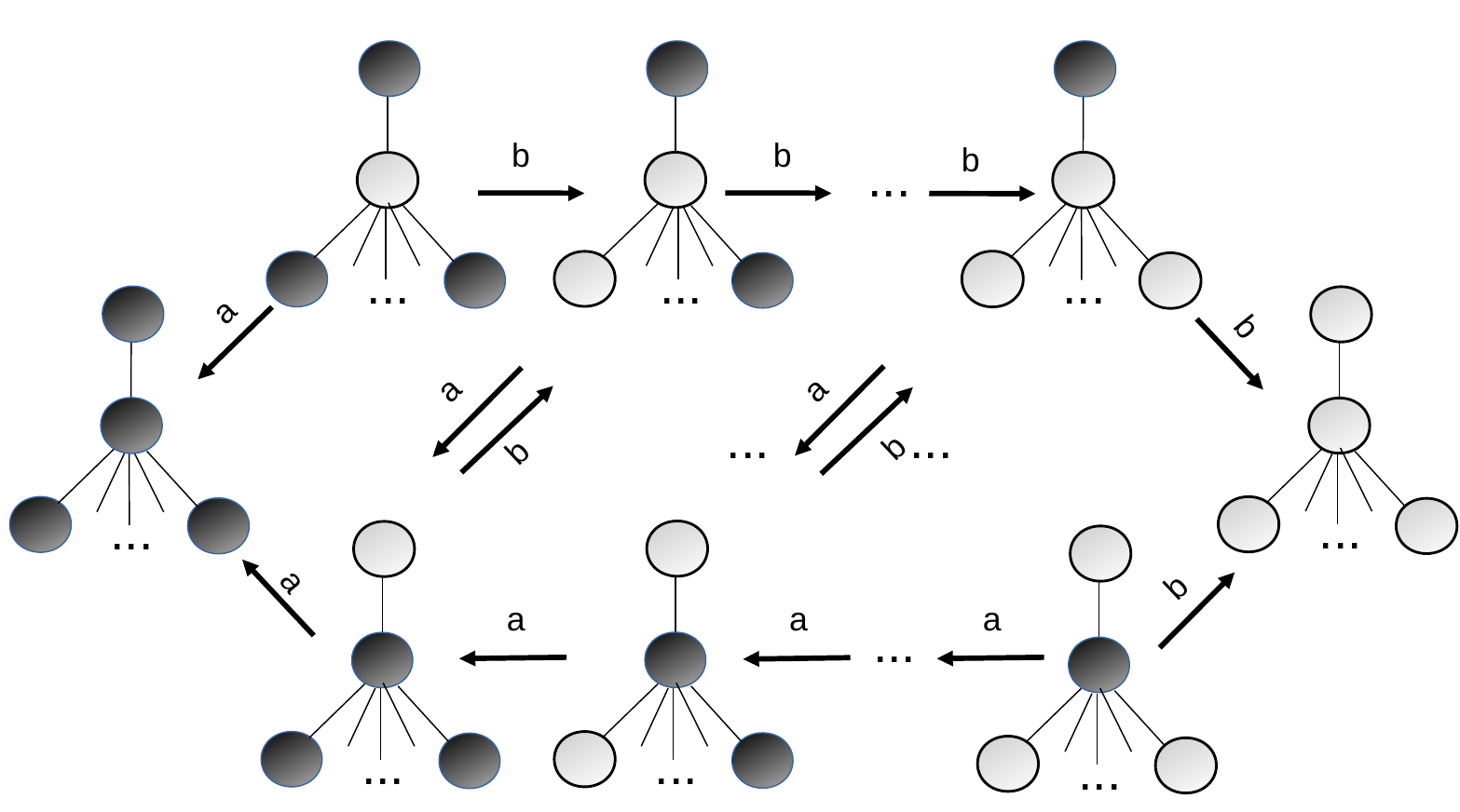}\centering 
\caption{{\bf Phase space of the HOpS model on a \emph{simple star}} with $k = 3$ 
leaves (top), arbitrary number of leaves (bottom). 
The arrows between each two states of the HOpS model
represent transition probabilities, which equal $a$ and $b$ correspondingly, 
Subsection \ref{anal_sol}. 
}
\label{variant}
\end{figure*}

Let us now consider a transformation matrix $P_k$ for the macro-states of the HOpS
where $k$ is the number of leaves of the simple star-like network. 
The size of matrix  $P_k$ is defined by the parametrized space $S^*$, 
as I showed in the case of the model on linear networks. 
Therefore a size of the transformation matrix $|P_k|$ 
for a star-like network is $ (2k+2) \times (2k+2)$, 
where $2k+2$ is the total number of states for the model on a star-network with $k$ leaves. 
This can be easily seen from the diagram Fig.~\ref{variant}. 
As an example,
I consider the \emph{HOpS model on a simple star network} 
for $k=3$ leaves. 
which has a transformation matrix  
 $|P_3| = 8\times8$. 
The structure of $P_3$ can be reordered in such a way that 
the first and the last matrix rows correspond to absorbing states of the system and 
rows  $i\in[2,2k+1]$ of the matrix $P_k$ correspond to transient states.
The elements of $P_3$ are rearranged in such a way  
that the nonzero matrix elements $a,b$ are parallel to the main diagonal. 
Then the matrix has a form: 
\begin{equation}
P_3 = \begin{pmatrix}
1 & 0 & 0 & 0 & 0 & 0 & 0 & 0\\
a & 0 & 0 & 0 & 0 & b & 0 & 0\\
0 & a & 0 & 0 & 0 & 0 & b & 0\\
0 & 0 & a & 0 & 0 & 0 & 0 & b\\
a & 0 & 0 & 0 & 0 & b & 0 & 0\\
0 & a & 0 & 0 & 0 & 0 & b & 0\\
0 & 0 & a & 0 & 0 & 0 & 0 & b\\
0 & 0 & 0 & 0 & 0 & 0 & 0 & 1\\
\end{pmatrix}
\label{tran_mat_star}
\end{equation}

In general, if the simple star-like network has $k$ leaves, 
the transformation matrix $P_k$ 
can be transformed to a Toeplitz matrix. 
This Toeplitz matrix has nonzero elements on four diagonals parallel to the main diagonal, 
which correspond to transition probabilities $a,b$ between states of the model: 
\begin{equation}
P_k = \begin{pmatrix}
1 & 0 & 0 & 0 & ... & 0 & 0 &...& 0 & 0\\
a & 0 & 0 & 0 & ... & b & 0 &...& 0 & 0\\
0 & a & 0 & 0 & ... & 0 & b &...&0 & 0\\
 & & & &... && &\\
a & 0 & 0 & 0 & ... & b & 0 &...& 0 & 0\\
0 & a & 0 & 0 & ... & 0 & b &...&0 & 0\\
 & & & &... && &\\
0 & 0 & 0 & 0 & ... & 0 & 0 &...&0 & 1\\
\end{pmatrix}
\end{equation}
where 
the distance between nonzero elements $a$ and $b$ elements in each row equals $k+1$. 
In analogy with the case of the HOpS model on a linear network 
knowing the transformation matrix $P_k$ of the system 
allows to find a stationary solution, an absorbing state of the model. 
It can be found as a vector state $Op(t\rightarrow \infty) $ for $  t \rightarrow \infty$ 
for the HOpS model on a star-like network 
by the evolution equation, Eq.~(\ref{evol_eq}): $Op(t+1) = Op(t) P_k$.
Then using geometric series we find 
$$Op(t\rightarrow \infty) = Op(0)(I-P_k)^{-1}$$
where $Op(0)$ 
is an initial state of the parameterized space $S^*$, $I$ is an identity matrix. 
\\
\\
\emph{Symmetric networks} have nontrivial group of automorphisms, 
i.e. it consists of more than from one element by the 
definition). 
This fact also causes that the invariant subspace of the full space of states 
is nontrivial, therefore a symmetry is emerged in the structure of the phase space. 
As a result, it is possible to describe network state using schematic diagrams, such as in Fig.~\ref{variant}. 
Although
in this section I considered \emph{linear and star-like networks}, 
some analytical results from Subsection \ref{anal_sol}  
are generalizable for solutions on the HOpS model dynamics on more general topologies. 
One example of such topology is a \emph{complex star} network   
with $k$ "long leaves", chains formed by $l$ nodes 
attached to a central node, Fig.~\ref{hierar} (b). 
Ultimately, 
presented methodological framework 
gives possibility to perform 
analysis of other DN models  
for network's topologies 
such as linear combination of star-networks trees, 
circular networks, fully connected graphs and others, which is discussed in the Outlook Section. 

\subsubsection{Consensus state for the HOpS model on a star networks}
Let us investigate the probability 
to reach the consensus state for the HOpS model on a star graph from the arbitrary initial condition. 
The HOpS model dynamics is described by the recursion formulas: 
\begin{equation}
W_n=\frac{1}{N+1}\,\frac{N-n}{N}\,B_n + \frac{N-n}{N+1}\,W_{n+1},  \\
B_n= \frac{1}{N+1}\,\frac{n}{N}\,W_n + \frac{n}{N+1}\,B_{n-1}  
\label{reccur}
\end{equation}
where $W_n$ denotes the probability for a model to be in a state with a white center, i.e. central node has opinion $0$,  
being surrounded by $n$ black nodes,   
while $B_n$ the probability for a model to be in a state with the black center, i.e. central opinion has opinion $1$,  
being surrounded by $n$ black nodes. 
Eqs.~(\ref{reccur}) are obtained by analogical observations to \cite{Krapivsky2003}, where 
the consensus time is estimated from the master equation for the majority rule model. 
Here we note that equations \ref{reccur} have  asymmetry, which leads to 
the asymmetric property of solutions. 
Eqs.~(\ref{reccur}) can be easily solved for small $N$. I i.e. for $N=2$ we get  %
$ W_0 = \tfrac{8}{35}, W_1 = \tfrac{12}{35}, W_2 = 1, 
B_0=0, B_1 = \tfrac{2}{35}, B_2 = \tfrac{13}{35}$.  
For bigger $N$ the transformation matrix method is needed.

\subsubsection{Analytical findings for the HOpS model on a star-like network}
To summarize, the main analytical findings for the HOpS model on a star-like network are: 
\\
(i) 
\emph{The HOpS model dynamics on star-like networks is described} by Propositions IV and V,  
the discrete phase space of the model is demonstrated in the schematic diagram, Fig.~\ref{variant}.  
As it can be seen from this digraph, transient states of the HOpS model form cycles. 
Moreover, since some random regular graphs are locally approximated as trees, 
then solution for the HOpS model dynamics on trees (or particularly, on star-like graphs) 
can give intuition for model solution on the whole graph. 
\\
(ii) 
\emph{The evolution of the HOpS model can be described 
using the transformation matrix framework.} An example of a transformation matrix equations for 
the HOpS model on a star network is illustrated in Eq.~(\ref{tran_mat_star}). 
\\
(iii) 
\emph{The general framework for studying a dynamical model on networks} 
can be formulated as the following. Firstly, one needs to 
find possible symmetries of the underlying network itself and to 
identify the invariant subspace of the full space of model states, as it has been demonstrated 
for a particular cases of networks in Proposition I. 
Then  
a phase space (or a part of a phase space) of a model can be represented as a digraph of SDSs, 
see Propositions V and VI. 
This would give qualitative characteristic of a model evolution.  
Secondly, one needs to estimate the transformation matrix as it was shown in Subsection \ref{hops_star}, 
which gives quantitative characteristic of a model evolution.

\subsection{Numerical results for the HOpS model on random networks}
\label{num_res}

The analytic solutions for DN models on special networks topologies 
were discussed in details in Section \ref{res_hops}. 
Further step is to get an intuition for the \emph{numerical results} of the HOpS model 
on random network topologies. 
Apart from inherent topological properties of random networks, 
these networks usually have a 
self-induced structure, which influences the flow 
of transport \cite{Holme2002}. This gives additional motivation 
to investigate random network models in the context of the HOpS model.

\subsubsection{The HOpS model dynamics on random Erd\H{o}s-Renyi networks}
\begin{figure*}[h]
\includegraphics[width=0.99 \textwidth]{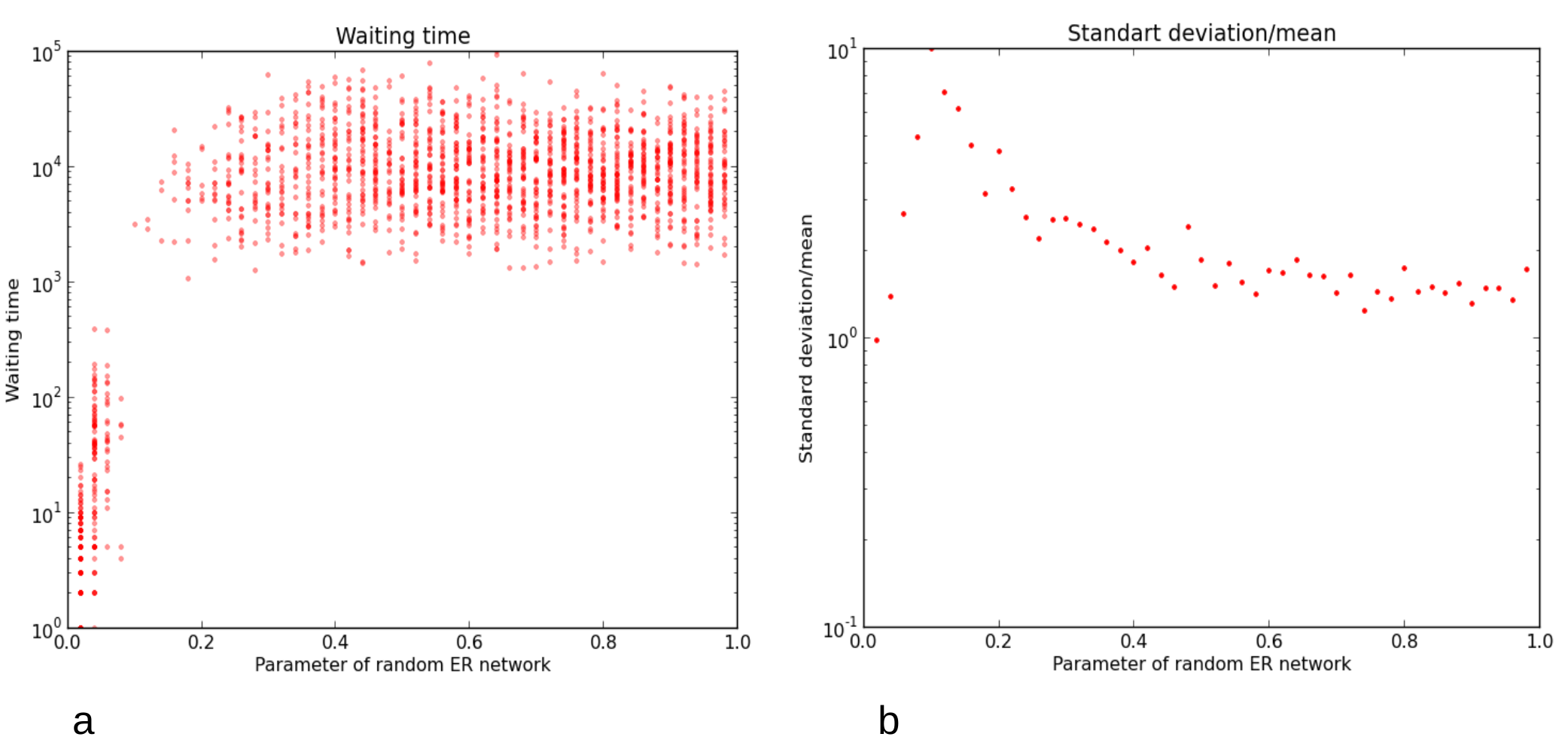}\centering 
\caption{
{\bf 
\emph{The waiting time} for the HOpS model 
on random Erd\H{o}s-Renyi networks  
depending on the probability of link creation $p$} is shown in subplot (a). 
Random realisations of the model 
are made for fixed parameters: $N=90$ nodes,  $T_{max} = 100000$, $\sigma =1$, 
with random statuses distribution for each model realisation. 
The waiting time (vertical axis) is calculated for each parameter $p$ (horisontal axis). 
\emph{Log-log} plot for the standard deviation/mean values 
of the waiting time is shown in subplot (b). A peak for a parameter of ER network $p\approx 0.15$ indicates the transition. }
\label{wait_time}
\end{figure*}

The random network type of the interest is 
random \emph{Erd\H{o}s-Renyi (ER) network} $G_{N,p}$ on $N$ nodes \cite{Erdos1959}.  
As it was already defined in Chapter II, 
each possible edge between two vertices 
is present in ER network with independent probability $p$, and absent with probability $1-p$.
It is important to mention 
that each realisation of ER network  with particular $p$ value 
is only a particular member of the entire statistical ensemble  
of random ER graphs. Hence, studying such model dynamics 
one needs to consider statistical ensemble 
of random networks instead of one particular random network realisation. 
Up till now in this chapter  
the space of model states of the HOpS model   
was mostly analyzed analytically. 
Underlying structure of ER networks is more complex than in "deterministic" symmetric networks, 
which leads to more intricate organisation of a phase space.  
I numerically examine behavior of the HOpS model 
by changing the model control parameters, 
specifically,  the ER network randomization parameter $p$. 
Then for each specific value of a control parameter $p$ I calculate the number of time steps 
till consensus is reached, hence 
I estimate so-called \emph{waiting} or \emph{relaxation time}. 
Notably, relaxation times for the HOpS model 
on symmetric deterministic networks with analytically estimated transformation matrix  $P$ 
can be estimated as a spectral gap of a matrix $P$, 
while waiting time for ER networks here is estimated numerically. 
\\
\\
\emph{The waiting time for the HOpS model on ER networks}. 
The waiting time as the function of an ER parameter $p$
for random configurations of network is shown in  
Fig.~\ref{wait_time}. The HOpS model dynamics 
is simulated on each realisation of $300$ ER networks $G(N,p)$ for $N=90$ nodes 
for each value $p$. 
The initial opinion distribution is randomized at each model run. 
The numerical simulations are performed 
until 
the HOpS model reaches a consensus. 
The waiting time in Fig.~\ref{wait_time} 
reaches the plateau 
for a value $p\approx 0.15$. In order to discuss the properties of the waiting time it is useful 
in parallel to examine the characteristics of ER networks. 
\\ 
Interestingly, ER networks have rich geometric properties, 
while being constructed by quite a simple rule. 
Let us consider the process of slow increase of ER parameter $p$. 
First, for low $p$ values  
ER network  $G_{N,p}$ consists from disconnected small components. 
Note that in each of such small component the consensus state of the HOpS in average is reached faster 
than on the whole fully connected network $G_{N,p}$ \cite{Krapivsky2013book}. 
Then if $p$ increases a newly introduced links start to appear more often between two disjoint clusters rather than between two nodes in the same cluster, 
and at some point \emph{the giant component} will be formed. For larger $p$ 
disconnected components of $G_{N,p}$ will form {a bigger predominantly tree-like network}, which still has a low link density, but spans   
more nodes than ER network for smaller $p$. 
Therefore the waiting time for the HOpS model on such tree-like network is expected to increase 
in comparison to the waiting time for the HOpS model on a tree-like component with lower link density. 
When $p$ continues to increase at some point ER network forms \emph{a fully connected component}. 
A fully connected component has high connectivity, 
hence the waiting time starts to decrease in comparison to 
the waiting time for smaller $p$ values.  
To summarize, for $p$ values larger than $p_c=1/N$, clusters in $G_{N,p}$ 
are small and tree-like \cite{Newman2003}. For $p > p_c$, a "giant" cluster is emerged, 
with all other components having size $O(\log N)$ \cite{sudakov2012}. 
Geometrically the percolation phenomenon in ER networks 
looks as the following: first for small $p$ there are just "seeds",
(small disconnected clusters of ER network $G(N,p)$).  
Then for larger $p$ these "seeds" emerge into a "forest" (a giant component), which is formed from cycles and trees linked together. 
The waiting time is affected by such reformation of 
components size distribution,   which is discussed in the following remark. 
\\
\\
\emph{Remark.}
The waiting time for the HOpS model dynamics on ER networks 
exhibits a transition in respect to the value of the ER parameter $p$. 
As it has been found numerically in Fig.~\ref{wait_time},  
the transition of the waiting time is observed for value $p\approx 0.15$. 
Some intuition for the analytical explanation to this observation is the 
following. 
Let us consider an ordinary diffusion process on a densely connected network \cite{Ghosh2011}. 
Let $T_d$ denote the waiting time till complete spreading through all nodes. 
Now let us compare $T_d$ 
with the waiting time $T_c$, which is the time of a complete spreading 
for a less densely connected network, which still forms a joined component,  
on the same set of nodes. 
Intuitively it is clear that the density of network connections influences   
the characteristics of dynamic process on networks, giving $T_d<T_c$,
if other parameters, such as  number of nodes in a network or network connectivity, 
are kept the same for both networks. 
In general, dynamics of DN models on arbitrary underlying structures is a challenging question, 
which is discussed further in the Conclusions and the Outlook Section.

\section{Conclusions and further directions}
The main conclusions of this chapter are enumerated below:
\begin{itemize}
\item 
\emph{1. 
A novel Heterogeneous OPinion-Status Model (HOpS) is introduced in this chapter. } 
The model dynamics on specific network topologies is described by Propositions I-VI.  
\item
\emph{
2. The HOpS model serves as a revealing test case for the new theoretical framework to  
describe a phase space of a discrete state model on networks}, 
Section \ref{res_hops}.  
The model setting links problems of discrete DN models with theory random walks \cite{Sokolov2011}. 
For symmetric networks a phase space of the model 
is regular, Figs.~\ref{diag_lin_net} and \ref{variant}. 
The approach to analyze a phase space of the HOpS model can be extended for more complex underlying network topologies using theory of sequential dynamical systems and the master equation approach \cite{Mortveit2013, Krapivsky2013book}.   
\item
\emph{
3. A theory of SDSs suggests a possible approach to study dynamics of DN models.} 
This can be significant both theoretically and practically
for a case of more general underlying graph structures. 
The novel approach to study DN models has its drawbacks, for instance, growing entanglement  of 
a phase space with increasing complexity of a topological structure of the underlying network.
\item 
 \emph{
4. The numerical results} for the model on Erd\H{o}s-Renyi (ER) random graphs 
are given in Subsection \ref{num_res}. 
The transition for a waiting time is observed for ER networks for 
an approximate value of ER randomization parameter $p \approx 0.15$, Fig.~\ref{wait_time}.
\item 
\emph{
5. The effects of heterogeneous spread 
of opinion on a network in the HOpS model were studied:  
particularly, it has been found 
that the speed of convergence towards the consensus state for the HOpS model on star-like graphs  
is affected by the heterogeneity parameter, introduced by a status distribution of nodes.} 
The network topology and the heterogeneity parameter are reflected in the HOpS model dynamics. 
As the result, the speed of convergence for the HOpS model on linear and star-like networks 
depends on the status difference parameter $\Delta_{st}$, Fig.~\ref{spectrum}. 
\\
Since HOpS model is a model of evolving opinion in heterogeneously distributed environment of statuses, 
one may ask, whether the results of the model dynamics  
are relevant for the real processes in society? 
The HOpS model dynamics  
can be an approximation of some complex systems, however the main motivation to study it was to 
deal with new types of dynamics and analysis methods. 
Most of the real world 
systems can be represented as complex DN models, 
therefore development of  general techniques to study such models 
is a topical issue in physics. 
\end{itemize}
\begin{figure*}[h]
\includegraphics[width=0.3 \textwidth]{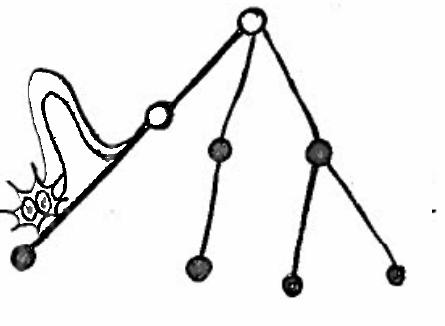} \centering
\caption{ {\bf Change of nodes states in the HOpS model on a tree-like network} 
can be presented as "random walkers", traveling on the network: 
at each time step one node has a probability to change its state. 
}
\label{rw_tree}
\end{figure*}

\subsubsection{Outlook}

The outlook includes open questions and generalizations of ideas from IV chapter: 
\\
1. \emph{What is the general 
method to find analytical solutions for the HOpS model on networks with arbitrary topology?} 
A method 
to analyze DN models, Section \ref{anal_sol}, 
has one particular limitation - the requirement of a regular  or 
acyclic structure of an underlying network, like a tree-like structure on Fig.~\ref{rw_tree}. 
However, the analytical solutions for the HOpS model can be extended in a quite straightforward way 
further for networks with more complex topologies. 
Also, in order to find the HOpS model solutions on asymmetric networks, 
adjacency matrices of asymmetric networks 
can be tackled as perturbed adjacency matrices of symmetric networks. 
It seems potentially interesting, to relate schematic diagrams, Fig.~\ref{variant}, with 
theory of group-reduced distributions \cite{Guichard} 
and 
the algebraic Galois theory for SDSs \cite{Lauben2001}. 
Additionally theory of Coxeter and Sylow's groups
could serve as promising instrument for theoretical description of SDSs \cite{Mortveit2013,macauley2008}, 
and also for analytical description of DN models. 
\\
\\
2. \emph{Which are further directions to study the HOpS model?} 
One potential generalization of the model
is to introduce the co-evolution of network topology and nodes' states. 
One of the possible model modification: 
let us call node $i$ an active node, if the random walk on the network 
is traveling on the network through this node. 
Additionally to this, 
some problems, such as to find analytical solutions for DN dynamics are closely interrelated to problems  
from the random walks theory \cite{nechaev2003, sinai1982, Sokolov2011}, 
particularly,  random walks on weighted directed networks and on circular networks with randomly added short-cuts.
Moreover, a novel heterogeneity parameter of the HOpS model can be introduced to any DN model,  
it induces an interesting complex dynamics, as it has been illustrated for linear and star-like networks. 
\\
\\
3. \emph{How the HOpS and voter models are interrelated?}
From the schematic diagram of the phase space of the model states of the HOpS model, Fig.\ref{variant},
it can be seen that in the HOpS model a node can change color from white to black even 
even if most of the nodes in its surrounding are white
(this is not possible in the voter model, where the mean opinion of the surrounding nodes defines the change of the opinion). 
These models 
can be described using general framework  \cite{Sokolov2011, Gleeson, ivaneyko, Redner2002}. 
\\
\\
4.
\emph{Which are possible applications of kinetic approach to the HOpS model?} 
There is a link between the HOpS model on a linear network from random initial conditions and the domain walls dynamics. 
The analytic solutions for the domain walls dynamics are known as a sum of Bessel functions \cite{Krapivsky2013book}.
A domain wall is simply a border between nodes with different opinions in the HOpS model. 
For instance, in Fig.~\ref{prob_line} there is one domain wall on a linear network, since 
there is only one edge between two sets of nodes with different opinions. 
In the language of domain walls it is clear that the HOpS model on a linear network with several domain walls 
has more complex dynamics rather than the HOpS model with only one domain wall. 
The homogeneous case, when a domain wall can move symmetrically with equal 
probabilities on a linear network, 
is also known as Sinai's random walk, or "quenched disorder" phenomenon \cite{Krapivsky1997, sinai1982}. 
\\
Furthermore, the methods, 
developed in Chapters II, III for evolving networks and flow-networks, can be applied  in order 
to quantify changes in generalizations of DN models on coevolving networks.

\bibliography{main}
\end{document}